\documentclass{aa}
\usepackage{natbib}
\newcount\longrefs
\def\adaa{\ifnum\longrefs=1 {Advances in Astronomy and Astrophysics}\else 
                           {Adv. Astron.\ Astrophys.}\fi}
\def\aap{\ifnum\longrefs=1 {Astron.\ Astrophys.}\else 
                           {A\hbox{\rm \&}A}\fi}
\def\aapr{\ifnum\longrefs=1 {Astron.\ Astrophys.\ Rev.}\else 
                            {A\hbox{\rm \&}AR}\fi}
\def\aaps{\ifnum\longrefs=1 {Astron.\ Astrophys.\ Suppl.}\else 
                            {A\hbox{\rm \&}AS}\fi}
\def\aj{\ifnum\longrefs=1 {Astron.\ J.}\else 
                          {AJ}\fi} 
\def\ao{\ifnum\longrefs=1 {Applied Optics}\else 
                           {Appl.\ Opt.}\fi} 
\def\aspcs{\ifnum\longrefs=1 {Astron.\ Soc.\ Pacific Conf. Series}\else 
                           {ASP Conf.\ Ser.}\fi} 
\def\apj{\ifnum\longrefs=1 {Astrophys.\ J.}\else 
                           {ApJ}\fi} 
\def\apjl{\ifnum\longrefs=1 {Astrophys.\ J. Lett.}\else 
                            {ApJ}\fi} 
\def\aplett{\ifnum\longrefs=1 {Astrophys.\ J. Lett.}\else 
                            {ApJ}\fi} 
\def\apjs{\ifnum\longrefs=1 {Astrophys.\ J. Suppl.}\else 
                            {ApJS}\fi}
\def\apss{\ifnum\longrefs=1 {Astrophys.\ and Space Science}\else 
                            {Ap\&SS}\fi}
\def\araa{\ifnum\longrefs=1 {Ann.\ Rev.\ Astron.\ Astrophys.}\else 
                            {ARA\hbox{\rm \&}A}\fi}
\def\azh{\ifnum\longrefs=1 {Astronomicheskii Zhurnal}\else 
                            {Astron.\ Zhur.}\fi}
\def\baas{\ifnum\longrefs=1 {Bull.\ Am.\ Astron.\ Soc.}\else 
                            {BAAS}\fi}
\def\bain{\ifnum\longrefs=1 {Bull.\ Astronom.\ Institutes Netherlands}\else
                            {Bull.\ Astr.\ Inst.\ Neth.}\fi}
\def\gca{\ifnum\longrefs=1 {Geochim.\ Cosmochim.\ Acta}\else 
                           {Geochim.\ Cosmochim.\ Acta}\fi}
\def\geo{\ifnum\longrefs=1 {Geophysical Journal}\else 
                           {Geophys.\ J.}\fi}
\def\grl{\ifnum\longrefs=1 {Geophys.\ Res.\ Lett.}\else 
                           {Geoph.\ Res.\ Lett.}\fi}
\def\iaucirc{\ifnum\longrefs=1 {IAU Circulars}\else 
                          {IAU Circ.}\fi}
\def\ip{\ifnum\longrefs=1 {in press}\else 
                          {in press}\fi}
\def\jgr{\ifnum\longrefs=1 {J.\ Geophys.\ Res.}\else 
                           {J.\ Geophys.\ Res.}\fi}  
\def\jrasc{\ifnum\longrefs=1 {J.\ Royal Astron.\ Soc.\ Canada}\else 
                           {JRAS Can.}\fi}  
\def\mnras{\ifnum\longrefs=1 {Mon.\ Not.\ Roy.\ Astron.\ Soc.}\else 
                             {MNRAS}\fi} 
\def\nat{\ifnum\longrefs=1 {Nature}\else 
                           {Nat}\fi}
\def\pasj{\ifnum\longrefs=1 {Pub.\ Astron.\ Soc.\ Japan}\else 
                            {PASJ}\fi} 
\def\pasp{\ifnum\longrefs=1 {Pub.\ Astron.\ Soc.\ Pacific}\else 
                            {PASP}\fi} 
\def\physscr{\ifnum\longrefs=1 {Physica Scripta}\else 
                            {Phys.\ Scrip.}\fi} 
\def\planss{\ifnum\longrefs=1 {Planetary \& Space Science}\else 
                            {Plan. \& Space Sci.}\fi} 
\def\procspie{\ifnum\longrefs=1 {Proc.\ SPIE}\else 
                            {Proc.\ SPIE}\fi} 
\def\qjras{\ifnum\longrefs=1 {Quarterly J.\ Royal Astron.\ Soc.}\else 
                            {QJRAS}\fi} 
\def\sa{\ifnum\longrefs=1 {Soviet Astron..}\else 
                               {Sov.\ Astron.}\fi}
\def\skytel{\ifnum\longrefs=1 {Sky \& Telescope}\else 
                            {Sky \& Tel.}\fi} 
\def\solphys{\ifnum\longrefs=1 {Solar Phys.}\else 
                               {Solar Phys.}\fi}
\def\ssr{\ifnum\longrefs=1 {Space Science Rev.}\else 
                               {Space\ Sci.\ Rev.}\fi}

\def\bibfiles{/home/bwillems/latex/bibtex/bibliofile}   

\def\aareferences{\longrefs=0  \bibliographystyle{/home/bwillems/latex/bibtex/aabib}
             \bibliography{/home/bwillems/latex/bibtex/aajour,\bibfiles}}

\def\dutch{\def\refname{Referenties}\def\abstractname{Samenvatting}%
  \def\bibname{Bibliografie}\def\chaptername{Hoofdstuk}%
  \def\appendixname{Bijlage}\def\contentsname{Inhoudsopgave}%
  \def\listfigurename{Lijst van figuren}\def\listtablename{Lijst van tabellen}%
  \def\indexname{Index}\def\figurename{Figuur}\def\tablename{Tabel}%
  \def\partname{Deel}\def\enclname{Bijlage(n)}\def\ccname{Ter attentie van}%
  \def\headtoname{Aan}\def\headpagename{Pagina}%
  \def\today{\number\day\space\ifcase\month\or januari\or februari\or maart\or%
     april\or mei\or juni\or juli\or augustus\or september\or oktober\or%
     november\or december\fi \space\number\year}%
  \typeout{
              >>>>> use hlatex209 for Dutch hyphenation <<<<< 
         }}
\newcounter{onefig} \newcounter{fignumber}
\newcount\nocaptions \newcount\nofigures \newcount\figwidth
\newcount\viewgraphs
  \def\paper{}  \def\figlabel{} 
\long\def\nextfig#1{\setcounter{figure}{\value{fignumber}}
  \addtocounter{fignumber}{1}
  \ifnum \viewgraphs=1 \newpage \pagestyle{empty} \fi 
  \ifnum\value{onefig}=0 #1 \fi                 
  \ifnum\value{onefig}=\value{fignumber} #1 \fi}
\def\figwidths#1#2{\ifnum \nocaptions=1 #2mm \else #1mm \fi}  
\def\paper#1{}  
\long\def\plotfig#1#2{\ifnum \nofigures=1 \else #2 \fi}
\long\def\captiontext#1{\ifnum \nofigures=1 \raggedright \fi 
   \ifnum \nocaptions=1 \paper
     \ifnum \viewgraphs=0 
       \newline  \mbox{}\hrulefill\mbox{} \newline 
       \newline label:~\{\figlabel\} 
     \fi 
     \else \ifnum \nofigures=0 \fi 
   #1 \fi}

\newcount\panelwidth \newcount\panelheight 
\newcount\bxmin \newcount\bymin \newcount\bxmax \newcount\bymax
\newcount\tbxmin \newcount\tbymin
\newcount\tpanelwidth \newcount\tpanelheight \newcount\tpdif
\def\panelsize #1,#2;{\panelwidth=#1 \panelheight=#2}  
\def\setbb #1,#2;#3,#4;#5,#6;{
  \tbxmin=#1 \tbymin=#2    
  \bxmin=#3 \bymin=#4      
  \bxmax=#5 \bymax=#6}     
\def\barepanel #1{%
  \ifnum\panelheight=0 
    \tpdif=\bymax \advance\tpdif by -\bymin
    \multiply \tpdif by \panelwidth
    \tpanelheight=\tpdif
    \tpdif=\bxmax \advance\tpdif by -\bxmin
    \divide \tpanelheight by \tpdif
  \else \tpanelheight=\panelheight \fi
  \epsfig{file=#1,%
     bbllx=\bxmin bp,bblly=\bymin bp,bburx=\bxmax bp,bbury=\bymax bp,clip=,%
     width=\panelwidth mm,height=\tpanelheight mm}}
\def\labelypanel #1{
  \ifnum\panelheight=0 
    \tpdif=\bymax \advance\tpdif by -\bymin
    \multiply \tpdif by \panelwidth
    \tpanelheight=\tpdif
    \tpdif=\bxmax \advance\tpdif by -\bxmin
    \divide \tpanelheight by \tpdif
  \else \tpanelheight=\panelheight \fi
  \tpdif=\bxmax \advance\tpdif by -\tbxmin
  \tpanelwidth=\panelwidth \multiply \tpanelwidth by \tpdif
  \tpdif=\bxmax \advance\tpdif by -\bxmin
  \divide \tpanelwidth by \tpdif
  \epsfig{file=#1,%
    bbllx=\tbxmin bp,bblly=\bymin bp,bburx=\bxmax bp,bbury=\bymax bp,%
    clip=,width=\tpanelwidth mm,height=\tpanelheight mm}}
\def\labelxpanel #1{%
  \ifnum\panelheight=0 
    \tpdif=\bymax \advance\tpdif by -\bymin
    \multiply \tpdif by \panelwidth
    \tpanelheight=\tpdif
    \tpdif=\bxmax \advance\tpdif by -\bxmin
    \divide \tpanelheight by \tpdif
  \else \tpanelheight=\panelheight \fi
  \tpdif=\bymax \advance\tpdif by -\tbymin
  \multiply \tpanelheight by \tpdif
  \tpdif=\bymax \advance\tpdif by -\bymin
  \divide \tpanelheight by \tpdif
  \epsfig{file=#1,%
    bbllx=\bxmin bp,bblly=\tbymin bp,bburx=\bxmax bp,bbury=\bymax bp,%
    clip=,width=\panelwidth mm,height=\tpanelheight mm}}
\def\labelxypanel #1{%
  \ifnum\panelheight=0 
    \tpdif=\bymax \advance\tpdif by -\bymin
    \multiply \tpdif by \panelwidth
    \tpanelheight=\tpdif
    \tpdif=\bxmax \advance\tpdif by -\bxmin
    \divide \tpanelheight by \tpdif
  \else \tpanelheight=\panelheight \fi
  \tpdif=\bxmax \advance\tpdif by -\tbxmin
  \tpanelwidth=\panelwidth \multiply \tpanelwidth by \tpdif
  \tpdif=\bxmax \advance\tpdif by -\bxmin
  \divide \tpanelwidth by \tpdif 
  \tpdif=\bymax \advance\tpdif by -\tbymin 
  \multiply \tpanelheight by \tpdif
  \tpdif=\bymax \advance\tpdif by -\bymin
  \divide \tpanelheight by \tpdif
  \epsfig{file=#1,%
    bbllx=\tbxmin bp,bblly=\tbymin bp,bburx=\bxmax bp,bbury=\bymax bp,%
    clip=,width=\tpanelwidth mm,height=\tpanelheight mm}}

\def\CC{\par \vspace*{-2ex} \footnotesize \baselineskip=8pt \begin{verbatim}}

\long\def\startignore #1\stopignore{}   

\def\setlistparams{         
  \topsep=0.7ex                 
  \itemsep=0.7ex                
  \leftmargini=3ex}             
\setlistparams                  

\newcounter{alistindex}       

\newcounter{romenumnr}

\newlength{\minipagewidth}

\newsavebox{\boxcontent}
\newcommand{\ovalhead}[1]{
  \unitlength=1cm
  \sbox{\boxcontent}{\mbox{~~{#1}~~}}
  \begin{center}
    \ifdim\wd\boxcontent>6ex 
    \ifdim\wd\boxcontent<8cm 
    \begin{picture}(8,3) \thicklines     
      \put(4.0,0.8){\oval(8,1.6)} 
      \put(0.0,0.7){\parbox{8cm}{
         \begin{center} \usebox{\boxcontent} \end{center}}}
    \end{picture}
    \else \ifdim\wd\boxcontent<12cm 
    \begin{picture}(12,3) \thicklines     
        \put(6.0,0.8){\oval(12,1.6)} 
        \put(0.0,0.7){\parbox{12cm}{
           \begin{center} \usebox{\boxcontent} \end{center}}}
    \end{picture}
    \else
    \begin{picture}(16,3) \thicklines     
        \put(8.0,0.8){\oval(16,1.6)} 
        \put(0.0,0.7){\parbox{16cm}{
           \begin{center} \usebox{\boxcontent} \end{center}}}
    \end{picture}
    \fi \fi \fi
  \end{center}} 

\newcounter{headnr}            
\newcounter{subheadnr}[headnr]
\newcounter{subsubheadnr}[subheadnr]
\def\head #1\par{
  \stepcounter{headnr}                          
  \vspace{2ex} \noindent                        
  {\bf \theheadnr~~~~#1}\\[1ex] \noindent}      
\def\subhead #1\par{  
  \stepcounter{subheadnr}
  \vspace{1.3ex} \noindent
  {\bf \theheadnr.\arabic{subheadnr}~~~#1}\\[0.3ex] \noindent}
\def\subsubhead #1\par{
  \stepcounter{subsubheadnr}
  \vspace{1.0ex} \noindent
  {\bf \theheadnr.\arabic{subheadnr}.\arabic{subsubheadnr}~~~#1}\\ \noindent}

\font\dropfont= cmr12 scaled \magstep5
\def\dropcap#1#2{{\noindent
    \setbox0\hbox{\dropfont #1}\setbox1\hbox{#2}\setbox2\hbox{(}%
    \count0=\ht0\advance\count0 by\dp0\count1\baselineskip
    \advance\count0 by-\ht1\advance\count0by\ht2
    \dimen1=.5ex\advance\count0by\dimen1\divide\count0 by\count1
    \advance\count0 by1\dimen0\wd0
    \advance\dimen0 by.25em\dimen1=\ht0\advance\dimen1 by-\ht1
    \global\hangindent\dimen0\global\hangafter-\count0
    \hskip-\dimen0\setbox0\hbox to\dimen0{\raise-\dimen1\box0\hss}%
    \dp0=0in\ht0=0in\box0}#2}

\def\level #1 #2#3#4{$#1 \: ^{#2} \mbox{#3} ^{#4}$}   

\def\mathstacksym#1#2#3#4#5{\def#1{\mathrel{\hbox to 0pt{\lower 
    #5\hbox{#3}\hss} \raise #4\hbox{#2}}}}

\mathstacksym\lta{$<$}{$\sim$}{1.5pt}{3.5pt} 
\mathstacksym\gta{$>$}{$\sim$}{1.5pt}{3.5pt} 
\mathstacksym\lrarrow{$\leftarrow$}{$\rightarrow$}{2pt}{1pt} 
\mathstacksym\lessgreat{$>$}{$<$}{3pt}{3pt} 

\usepackage{graphics}
\def\bbbr{{\rm I\!R}}

\begin{document}
\title{Nonadiabatic resonant dynamic tides and orbital evolution in
close binaries}
\author{B.\ Willems\inst{1,}\thanks{\emph{Present address:} Department of
    Physics and Astronomy, Open University, Walton Hall, Milton
    Keynes, MK7 6AA, UK}
    \and T.\ Van Hoolst\inst{2,1}
    \and P.\ Smeyers\inst{1}}
\institute{Instituut voor Sterrenkunde, Katholieke Universiteit
Leuven, Celestijnenlaan 200\,B, B-3001 Heverlee, Belgium \and
Koninklijke Sterrenwacht van Belgi\"e,
  Ringlaan 3, B-1180 Brussel, Belgium}
\date{Received date; accepted date}
\offprints{B.Willems@open.ac.uk}


\abstract{
This investigation is devoted to the effects of nonadiabatic
resonant dynamic tides generated in a uniformly rotating stellar
component of a close binary. The companion is considered to move in a
fixed Keplerian orbit, and the effects of the centrifugal force and the
Coriolis force are neglected. Semi-analytical solutions for the
linear, nonadiabatic resonant dynamic tides are derived by means of a
two-time variable expansion procedure. The solution at the lowest
order of approximation consists of the resonantly excited oscillation
mode and displays a phase shift with respect to the tide-generating
potential. Expressions are established for the secular variations of
the semi-major axis, the orbital eccentricity, and the star's angular
velocity of rotation caused by the phase shift. The orders of
magnitude of these secular variations are considerably larger than
those derived earlier by \citet{Zahn1977} for the limiting case of
dynamic tides with small frequencies. For a $5\,M_\odot$ ZAMS star, an
orbital eccentricity $e = 0.5$, and orbital periods in the range from
2 to 5 days, numerous resonances of dynamic tides with second-degree
lower-order $g^+$-modes are seen to induce secular variations of the
semi-major axis, the orbital eccentricity, and the star's angular
velocity of rotation with time scales shorter than the star's nuclear
life time.
}

\maketitle

\section{Introduction}

Various investigations have already been devoted to tidal actions
in components of close binaries and their effects on orbital
evolution.

A commonly used approach for the study of dynamic tides
rests on a Fourier decomposition of
the tide-generating potential in terms of multiples of the
orbital frequency. For each forcing frequency, the dynamic tide is
then determined by integration of the system of differential
equations governing linear, forced oscillations of an equilibrium
star on the assumption of a fixed orbit.

\citet{Zahn1975} used this approach in order to determine the influence
of the radiative damping in the nonadiabatic surface layers of a
spherically symmetric star on low-frequency dynamic tides. He
considered a star consisting of a convective core and a radiative
envelope. In developing an asymptotic theory, he distinguished
between the interior layers, where the forced oscillations are nearly
isentropic, and a boundary layer, where the radiative damping
is important.

Because of the radiative dissipation near the star's surface,
dynamic tides do not have the same properties of symmetry as the
exciting potential, so that the companion exerts a torque on the
star which may tend to synchronise the star's rotation with the
orbital motion. According to Zahn, the torque appears to be strong
enough to achieve synchronisation, for relatively close binaries,
in a time short in comparison with the star's nuclear life time.
Consequently, Zahn considered dynamic tides with radiative
dissipation as an efficient process for the synchronisation of
components of close binaries that do not have a convective
envelope.

A generalisation of Zahn's treatment to include the effects of the
Coriolis force in the case of a slowly rotating component was
given by \citet{Roc1989}.

The system of equations governing the non-adiabatic tidal response
of a spherically symmetric star was integrated numerically by
\citet{SP1983}. For simplicity, the authors
restricted themselves to small orbital eccentricities and
neglected the perturbation of the gravitational potential caused
by the tidal response. They presented results for main
sequence, non-rotating tidally distorted stars and concluded that
the ``the spin up of the massive component appears to be rather
inefficient \ldots, unless the system could be locked into
resonance \ldots''.
Subsequently, \citet{SP1984} determin\-ed the tidal response for an
evolved stellar model of $20\, M_\odot$ and found a substantial
increase of the effectiveness of tidal evolution during the late
phases of core hydrogen burning.

On his side, \citet{Tas1987} argued that the general trend toward
synchronism that is found in early-type binaries, cannot be
explained solely by Zahn's theory on the action of radiative
damping on dynamic tides. He presented a hydrodynamical mechanism
involving large-scale, mechanically driven currents which
transport angular momentum.

More recently, the work of Savonije \& Papaloizou was
extended by a series of investigations on uniformly rotating
spherically symmetric stars in which the effects of the Coriolis force
were taken into account \citep{SPA1995,PS1997,SP1997,Lai1997,Wit1999a}.

The effect of resonance locking on the tidal evolution of massive
binaries was studied by \citet{Wit1999b,Wit2001}.
The authors concluded that when both orbital and stellar evolution are
taken into account, a dynamic tide can easily become locked in a
resonance for a prolonged period of time. They furthermore found that,
during a resonance, a
significant enhancement of the secular evolution of the semi-major
axis and the orbital eccentricity occurs.

In this study, our aim is first to describe the influence of the
perturbation of the radiative energy transfer on
tidal resonances with free oscillation modes in components of
close binaries. To this end, we use a two-time variable expansion
procedure leading to semi-analytical solutions for the resonant
dynamic tides and the resulting perturbation of the star's
external gravitational field. This investigation is an extension of
an earlier investigation of \citet{SWV1998} (hereafter
referred to as Paper~I), in which the resonant dynamic tides were
treated in the isentropic approximation.

In the second part of our investigation, we determine the effects
of the nonadiabaticity of the tidal motions on the orbital
evolution of a close binary in cases of resonances of dynamic
tides with free oscillation modes $g^+$.

We consider a uniformly rotating component of a close binary but we
still neglect the effects of the centrifugal force and the Coriolis
force on the resonant dynamic tides.

The plan of the paper is as follows. In Sect.~2, we present the basic
assumptions and the equations governing nonadiabatic forced
oscillations of a rotating spherically symmetric star. In Sect.~3, the
tide-generating potential is expanded in terms of spherical harmonics
and in Fourier series in terms of the companion's mean motion. In
Sect.~4, we derive the lowest-order solutions for the nonadiabatic
resonant dynamic tides.  Sect.~5 is devoted to the derivation of the
equations governing the secular changes of the semi-major axis, the
orbital eccentricity, and the longitude of the periastron due to the
perturbation of the star's external gravitational field. In Sect.~6,
we determine the torque exerted by the companion on the tidally
distorted star and the associated rate of secular change of the star's
rotational angular velocity.  In Sect.~7, we present phase shifts for
resonant dynamic tides in a $5\,M_\odot$ zero-age main sequence
star. We also consider time scales of orbital evolution for short
orbital periods ranging from 2 to 5 days and for the orbital
eccentricity $e = 0.5$. Finally, Sect.~8 is devoted to concluding
remarks.

\section{Basic assumptions and governing equations}

Consider a close binary system of stars with masses $M_1$ and $M_2$
that are orbiting around each other under the influence of their mutual
gravitational force. The first star is assumed to rotate uniformly
around an axis perpendicular to the orbital plane in the sense of
the orbital motion, and the second star, called the companion,
is considered to be a point
mass describing a {\it fixed} Keplerian orbit around the star.

As in Paper~I, we describe the oscillations of the star with respect
to a corotating frame of reference $C_1 x^{\prime \prime 1}
x^{\prime \prime 2}
x^{\prime \prime 3}$ whose origin coincides with the mass centre $C_1$
of the star and whose $x^{\prime \prime 1} x^{\prime \prime2}$-plane
coincides with the orbital plane of the binary. The direction of the
$x^{\prime \prime 3}$-axis corresponds to the direction of the star's
angular velocity $\vec{\Omega}$. With respect
to this corotating frame of reference, we introduce a system of
spherical coordinates $\vec{r}=(r,\theta,\phi)$. We consider the
spherical coordinates $r$, $\theta$, $\phi$ as generalised coordinates
$q^1$, $q^2$, $q^3$.

The tidal force generated by the companion is introduced as a small
time-dependent force which perturbs the hydrostatic equilibrium of a
spherically symmetric static star. Correspondingly, the tide generated
by the companion is considered as a {\it forced} perturbation of the
star's equilibrium.

Let $\varepsilon_T W\left(\vec{r},t\right)$ be the tide-generating
potential, where $\varepsilon_T$ is a small dimensionless parameter
defined as
\begin{equation}
\varepsilon_T = \left({ R_1\over a}\right)^3\, {M_2\over M_1}.
   \label{b:3}
\end{equation}
Here $R_1$ is the radius of the spherically symmetric equilibrium
star, and $a$ the semi-major axis of the companion's relative
orbit.

Furthermore, let $\left(\delta q^j\right)_T$, with $j = 1, 2, 3,$ be
the contravariant components
of the tidal displacement in the star with respect to the local
coordinate basis $\partial/\partial q^1$, $\partial/\partial q^2$,
$\partial/\partial q^3$. If one neglects the effects of the
centrifugal force and the Coriolis force, these components are
governed by the equations
\begin{equation}
g_{ij}\, {{\partial^2 \left(\delta q^j\right)_T}\over {\partial t^2}}
  + U_{ij} \left(\delta q^j\right)_T = - \varepsilon_T\,
  {{\partial W}\over {\partial q^i}}, \, \, \, i = 1,2,3.   \label{b:1}
\end{equation}
The $g_{ij}$ are the components of the metric tensor, and the
$U_{ij}$ the tensorial operators applying to free linear,
isentropic oscillations of a spherically symmetric star, which are
defined by
\begin{equation}
U_{ij} \left({\delta q^j}\right)_T
  = {{\partial \left(\delta \Phi\right)_T} \over {\partial q^i}}
  - {{\left(\delta \rho\right)_T} \over {\rho^2}}\,
  {{\partial P} \over {\partial q^i}} + {1\over \rho}\,
  {{\partial \left(\delta P\right)_T}\over {\partial q^i}}.
  \label{b:2}
\end{equation}
In these equations, $P$ is the pressure, $\rho$ the mass density,
and $\Phi$ the gravitational potential. When the operator $\delta$ is
applied to a quantity, the Lagrangian perturbation of that quantity is
taken.

The Lagrangian perturbations of the mass density and the gravitational
potential are determined by the equation expressing the conservation
of mass
\begin{equation}
{{\left(\delta \rho \right)_T } \over \rho}
  = - \nabla_j \left(\delta q^j\right)_T  \label{la:15}
\end{equation}
and by Poisson's integral equation
\begin{eqnarray}
\lefteqn{\left(\delta \Phi\right)_T =
  \left(\delta q^j\right)_T \nabla_j \Phi }  \nonumber \\
 & & - G \int_V \rho \left(\vec{r^\prime}\right)
  \left(\delta q^{\prime j}\right)_T \left(\vec{r^\prime}\right)
  \left[ \nabla_{\prime j} |\vec{r}-\vec{r^\prime}|^{-1} \right]
  dV \left(\vec{r^\prime}\right).  \label{la:16}
\end{eqnarray}
In the latter equation, $G$ is the gravitational constant, and $V$
the volume of the
spherically symmetric equilibrium star. The operators $\nabla_j$ and
$\nabla_{\prime j}$ are the operators of partial differentiation with
respect to the generalised coordinates $q^j$ and $q^{\prime j}$
when they are applied to a scalar quantity, and are the
operators of covariant differentiation when they are applied to a
vector or a tensor component.

For tides considered as linear, {\it nonadiabatic} forced
perturbations of the spherically symmetric equilibrium star, the
Lagrangian perturbation of the pressure is
determined by the equation for the rate of change of thermal energy
\begin{equation}
{\partial \over {\partial t}} \left[
  {{\left( \delta P \right)_T} \over P} - \Gamma_1
  {{\left( \delta \rho \right)_T} \over \rho} \right]
  = {{\left( \Gamma_3 - 1 \right) \rho} \over P}\,
  \left( \delta\, {{dQ} \over {dt}} \right)_T.  \label{b:7}
\end{equation}
Here $\Gamma_1 \equiv (\partial \ln P/\partial \ln \rho)_S$ and
$\Gamma_3 - 1 \equiv (\partial \ln T/\partial \ln \rho)_S$ are
generalised isentropic coefficients, $T$ is the temperature,
and $dQ/dt$ is the rate of energy exchange per unit time and unit
mass. In the regions of the star where the energy transfer is purely
radiative, $dQ/dt$ is given by
\begin{equation}
{{dQ} \over {dt}} = \varepsilon_1 - {1 \over \rho}\, \nabla \cdot
  \vec{F}_{\rm R}, \label{b:7b}
\end{equation}
where $\varepsilon_1$ is the amount of energy generated by the
thermonuclear reactions per unit time and unit mass, and
$\vec{F}_{\rm R}$ the radiative energy flux.

We make the equations dimensionless by expressing the time $t$, the
radial coordinate $r$, the mass density $\rho$, the pressure $P$, both
the gravitational potential $\Phi$ and the tide-generating potential
$\varepsilon_T\,W$, and the rate of energy exchange per unit mass $dQ/dt$
respectively in the units $\left[R_1^3/(GM_1)\right]^{1/2}$, $R_1$,
$M_1/\left(4 \pi R_1^3 \right)$, $GM_1^2/\left(4 \pi R_1^4 \right)$,
\linebreak $GM_1/R_1$, and
$L_1/M_1$, where $L_1$ is the luminosity of the star.

Furthermore, we express the Lagrangian perturbation of the rate of energy
exchange during the tidal motions formally in terms of the components
of the tidal displacement as
\begin{equation}
\left(\delta\, {{dQ}\over{dt}}\right)_T =
  {\cal V}_j \left(\delta q^j \right)_T,  \label{b:8}
\end{equation}
where the ${\cal V}_j$ are linear operators. Eq.\ (\ref{b:7}) for
the rate of change of thermal energy can then be rewritten as
\begin{equation}
{\partial \over {\partial t}}\! \left[
  {{\left( \delta P \right)_T} \over P} + \Gamma_1
  \nabla_j\! \left(\delta q^j\right)_T \right]\!
  = C\, {{\left( \Gamma_3 - 1 \right) \rho} \over P}\,
  {\cal V}_j \left(\delta q^j \right)_T,  \label{b:9}
\end{equation}
where
\begin{equation}
C \equiv \left( {R_1^3 \over {G\, M_1}} \right)^{1/2} /
  \left( {{G\, M_1^2} \over {R_1\, L_1}} \right)  \label{b:10}
\end{equation}
is the ratio of the dynamic time scale to the Helm\-holtz-Kelvin time
scale of the star.

\section{The tide-generating potential}

Following \citet{Pol1990}, we expand the
tide-generating potential in terms of unnormalised spherical
harmonics $Y_\ell^m(\theta,\phi)$ and in Fourier series in terms
of multiples of the companion's mean motion $n$. The expansion
takes the form
\begin{eqnarray}
\lefteqn{ \varepsilon_T\, W\left(\vec{r},t\right)
  = \varepsilon_T\, \sum_{\ell=2}^4 \sum_{m=-\ell}^\ell
  \sum_{k=-\infty}^\infty W_{\ell,m,k}\left(\vec{r}\,\right) }
  \nonumber \\
 & & \exp \left[{\rm i} \left(\sigma_T\, t
  - k\, n\, \tau \right)\right],  \hspace{3.0 cm}  \label{b:4}
\end{eqnarray}
where
\begin{equation}
W_{\ell,m,k}\left(\vec{r}\,\right) = -
  c_{\ell,m,k}\, \left({r\over R_1}\right)^\ell\,
  Y_\ell^m(\theta,\phi),   \label{b:5}
\end{equation}
$\sigma_T$ is
the forcing angular frequency with respect to the corotating frame
of reference determined as
\begin{equation}
\sigma_T = k\, n + m\, \Omega,  \label{b:12}
\end{equation}
and $\tau$ is a time of periastron passage. The Fourier
coefficients $c_{\ell,m,k}$ are defined as
\begin{eqnarray}
\lefteqn{c_{\ell,m,k} = \displaystyle
  {{(\ell-|m|)!} \over {(\ell+|m|)!}}\, P_\ell^{|m|}(0)
  \left({R_1\over a}\right)^{\ell-2}
  {1\over {\left({1 - e^2}\right)^{\ell - 1/2}}} } \nonumber \\
 & & {1\over \pi} {\int_0^\pi (1 + e\, \cos v)^{\ell-1}\,
  \cos (k\, M + m\, v)\, dv}. \label{pot:2}
\end{eqnarray}
In this definition, $P_\ell^{|m|}(x)$ is an associated Legendre polynomial
of the first kind, and $M$ and $v$ are respectively the mean and the true
anomaly of the companion in its relative orbit.

>From Definition (\ref{pot:2}), it follows that the Fourier
coefficients $c_{\ell,m,k}$ depend on the orbital eccentricity.
They decrease as the multiple $k$ of the mean motion increases.
The decrease is slower for higher orbital eccentricities.
Consequently, the number of terms that has to be taken into
account in Expansion (\ref{b:4}) of the tide-generating
potential increases with increasing values of the orbital
eccentricity. For illustration, the logarithms of the absolute
values of the second-degree Fourier coefficients $c_{2,m,k}$, with
$m=-2,0,2$, are displayed in Fig.~\ref{c2mk} as functions of $k$
for the orbital eccentricities $e = 0.25$ and $e = 0.5$.

\begin{figure}
\resizebox{\hsize}{!}{\includegraphics{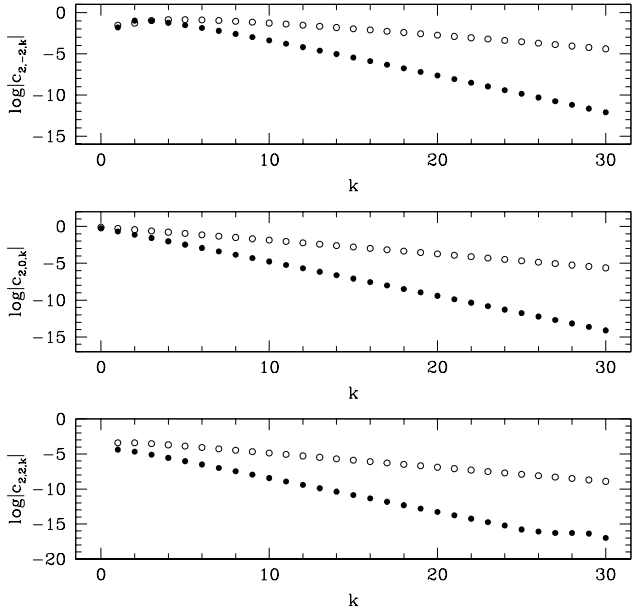}}
\caption{Logarithmic representation of
the absolute values of the Fourier coefficients $c_{2,m,k}$, with
$m=-2,0,2$, as functions of $k$, for the orbital
eccentricities $e=0.25$ (solid circles) and $e=0.5$ (open circles).}
\label{c2mk}
\end{figure}

\section{Lowest-order solutions for nonadiabatic resonant dynamic tides}
\label{loword}

We consider a single partial dynamic tide generated by the term
\begin{equation}
\varepsilon_T\, W_{\ell,m,k}\left(\vec{r}\,\right) \, \exp \left[
  {\rm i} \left( \sigma_T\, t -k\, n\, \tau \right) \right]
  \label{b:11}
\end{equation}
of Expansion (\ref{b:4}) of the tide-generating potential, and assume
that its forcing angular frequency $\sigma_T$ is close to the degenerate
eigenfrequency $\sigma_{\ell, N}$ of the star's free linear,
isentropic oscillation modes of radial order $N$ that
belong to the spherical harmonics $Y_\ell^m(\theta,\phi)$ of degree
$\ell$ and the various admissible azimuthal numbers $m$. We define
the relative frequency difference $\varepsilon$ as
\begin{equation}
\varepsilon = {{\sigma_{\ell,N} - \sigma_T}\over {\sigma_{\ell,N}}}
  \label{b:15}
\end{equation}
and assume $\varepsilon$ to be of the order of $\varepsilon_T$:
\begin{equation}
\varepsilon = f\, \varepsilon_T, \,\,\, f \in \bbbr.  \label{b:15b}
\end{equation}

In addition, we consider the nonadiabatic effects, which are rendered
by the right-hand member of Eq. (\ref{b:9}), to be of the order of
$\varepsilon_T$ at least in some region near the star's
surface and, for convenience, set
\begin{equation}
C = C^\prime\, \varepsilon_T, \,\,\, C^\prime\, \in \bbbr.
  \label{b:15c}
\end{equation}

We determine the combined effects of the resonance of the dynamic tide
and the nonadiabaticity of the tidal motions by means of an
appropriate two-time variable expansion procedure (\citealt{KC1981},
Section 3.2.7; \citeyear{KC1996}, Section 4.3.1). In this expansion
procedure, the non\-adiabatic terms are evaluated in the
quasi-isentropic approximation. A similar perturbation procedure was
used by \citet{Ter1998} for the evaluation of the nonadiabaticity of
tidal motions.

In view of the application of the two-time variable expansion
procedure, we pass on to a new dimensionless time variable
\begin{equation}
t^\ast = \sigma_{\ell,N}\, t.  \label{b:14}
\end{equation}
Adopting $\varepsilon_T$ as the small expansion parameter, we
introduce a fast and a slow time variable as
\begin{equation}
t^+ = t^\ast  \left[1+O\left(\varepsilon_T^2\right) \right]
  \label{b:18}
\end{equation}
and
\begin{equation}
\tilde{t} = \varepsilon_T\, t^\ast,  \label{b:19}
\end{equation}
and expand the components of the tidal displacement in terms of
the star's linear, isentropic oscillation modes $s^\prime$ as
\begin{equation}
\left(\delta q^j\right)_T \left(\vec{r},t\right) =
  \sum_{s^\prime} \Bigl[ \sum_{n=0}^\infty \varepsilon_T^n
  F_{s^\prime}^{(n)}\left(t^+\!,\tilde{t}\,\right) \Bigr]
  \left(\delta q^j\right)_{s^\prime}\!\left(\vec{r}\right),
  \label{b:20}
\end{equation}
where $j = 1,2,3$.
The expansions can be restricted to the star's sphero\-idal modes,
since neither the tidal force nor the nonadiabatic effects produce any
vorticity around the normal to the local spherical equipotential
surface of the equilibrium star. A given spheroidal mode $s^\prime$ is
characterised by the degree $\lambda^\prime$ and the azimuthal number
$\mu^\prime$ of the spherical harmonic to which it belongs, and by the
radial order $\nu^\prime$. The components of the Lagrangian
displacement and the angular eigenfrequency of the oscillation mode
$s^\prime$ obey the wave equations
\begin{equation}
\sigma^2_{\lambda^\prime \!,\nu^\prime}\, g_{ij}
  \left({\delta q^j}\right)_{\lambda^\prime \!,\mu^\prime \!,\nu^\prime}
  \! - U_{ij}
  \left({\delta q^j}\right)_{\lambda^\prime \!,\mu^\prime \!,\nu^\prime}
  \! = 0, \,\,\, i=1,2,3.   \label{b:13}
\end{equation}
We also introduce an expansion for the Lagrangian perturbation of the
pressure as
\begin{equation}
\left( \delta P \right)_T \left(\vec{r},t\right) =
  \sum_{n=0}^\infty \varepsilon_T^n \left( \delta P \right)_T^{(n)}
  \left(\vec{r},t^+\!,\tilde{t}\,\right).  \label{b:13b}
\end{equation}

Substitution of Expansions (\ref{b:20}) and (\ref{b:13b}) into Eq.\
(\ref{b:9}), application of the chain rule, and integration with
respect to the fast time variable yield, at order $\varepsilon_T^0$,
\begin{equation}
{{\left( \delta P \right)_T^{(0)}} \over P} = - \Gamma_1\,
  \sum_{s^\prime} F_{s^\prime}^{(0)}\left(t^+\!, \tilde{t}\,\right)
  \nabla_j \left(\delta q^j\right)_{s^\prime}.  \label{b:22}
\end{equation}

Introduction of the latter expansion into Eqs.\ (\ref{b:1}) leads to the
equations
\begin{eqnarray}
\lefteqn{ g_{ij}\, {\partial^2\over {\partial t^{+ 2}}}
  \sum_{s^\prime} F_{s^\prime}^{(0)}\left(t^+\!, \tilde{t}\,\right)\!
  \left(\delta q^j\right)_{s^\prime} }
  \nonumber \\
 & & + {1\over {\sigma^2_{\ell,N}}} \sum_{s^\prime}
  F_{s^\prime}^{(0)}\left(t^+\!,\tilde{t}\,\right) U_{ij}\!
  \left(\delta q^j\right)_{s^\prime} = 0, \,\,\, i=1, 2, 3.  \label{b:23}
\end{eqnarray}
We transform these equations by using wave Eqs.\ (\ref{b:13}) and
multiplying all terms by $\rho\, \overline{ \left({\delta
q^i}\right)_s }$, where the bar denotes the
complex conjugate, and $s$ stands for the degree $\lambda$, the azimuthal
number $\mu$, and the radial order $\nu$ of a given
spheroidal mode. Next, we integrate over the volume
of the spherically symmetric equilibrium star and take into account
the orthogonality property of the spheroidal modes. The resulting
equation for the functions $F_s^{(0)}\left(t^+\!, \tilde{t}\,\right)$
takes the form
\begin{equation}
{\partial^2\over {\partial t^{+2}}}\, F_s^{(0)}\left(t^+\!,
  \tilde{t}\,\right) + {\sigma^2_{\lambda, \nu} \over \sigma^2_{\ell,N}}\,
  F_s^{(0)}\left(t^+\!, \tilde{t}\,\right) = 0.
  \label{b:24}
\end{equation}
A general solution is given by
\begin{equation}
F^{(0)}_s\left(t^+\!, \tilde{t}\,\right) =
  A^{(0)}_s\left(\tilde{t}\,\right) \exp \left( \mbox{i}
  {\sigma_{\lambda, \nu} \over \sigma_{\ell,N}}\, t^+ \right),
   \label{b:25}
\end{equation}
where $A^{(0)}_s\left(\tilde{t}\,\right)$ is a yet undetermined complex
function of the slow time variable $\tilde{t}$.

At order $\varepsilon_T$, substitution of Solutions (\ref{b:25}) into
Eq.\ (\ref{b:9}) and integration with respect to the fast time
variable yield
\begin{eqnarray}
\lefteqn{ {{\left( \delta P \right)_T^{(1)}} \over P} = - \Gamma_1
  \sum_{s^\prime} \bigg\{
  F_{s^\prime}^{(1)}\!\left(t^+\!, \tilde{t}\,\right) \nabla_j\!
  - {\rm i}\, C^\prime\,
  {{\left( \Gamma_3 - 1 \right) \rho} \over P} } \nonumber \\
 & & A^{(0)}_{s^\prime}\left(\tilde{t}\,\right)
  \sigma_{\lambda^\prime, \nu^\prime}^{-1}
  \left[ \exp \left( {\rm i}\, {\sigma_{\lambda^\prime, \nu^\prime}
  \over \sigma_{\ell,N}}\, t^+ \right) \right]
  {\cal V}_j \bigg\}
  \left(\delta q^j\right)_{s^\prime}.  \label{b:27}
\end{eqnarray}
In this equality, the second term in the right-hand member
is the contribution of the nonadiabatic effects to the
Lagrangian perturbation of the pressure.

By proceeding in a similar way as for the derivation of Eq.\
(\ref{b:24}), it follows that the functions $F_s^{(1)}\left(t^+\!,
  \tilde{t}\,\right)$ are governed by the inhomogeneous equation
\begin{eqnarray}
\lefteqn{ {\partial^2\over {\partial t^{+2}}}\, F_s^{(1)}\left(t^+\!,
  \tilde{t}\,\right) + {\sigma^2_{\lambda, \nu} \over \sigma^2_{\ell,N}}\,
  F_s^{(1)}\left(t^+\!, \tilde{t}\,\right) }  \nonumber \\
 & & = - 2\, {\rm i}\, {{dA^{(0)}_s}\over {d\tilde{t}}}\,
  {\sigma_{\lambda,\nu} \over \sigma_{\ell,N}}\,
  \exp \left( {\rm i}\,{\sigma_{\lambda,\nu} \over \sigma_{\ell,N}}\,
   t^+ \right)  \nonumber \\
 & & -\, {1 \over {\sigma_{\ell,N}^2\, {\cal N}_s}}\,
  \left[ \int_V \rho\, {{\partial W_{\ell,m,k}}\over {\partial q^i}}\,
  \overline{\left({\delta q^i}\right)_s}\, dV \right]  \nonumber \\
 & & \;\;\; \Big\{\exp \left[- {\rm i} \left(f\, \tilde{t}
  + k\, n\, \tau \right)\, \right] \Big\}
  \exp \left( {\rm i} t^+ \right)  \nonumber \\
 & & +\, {{{\rm i}\, C^\prime} \over
  {\sigma_{\ell,N}^2\,{\cal N}_s}}\! \sum_{s^\prime}
  \bigg\{\! \int_V \overline{\left( \delta q^i \right)_s}\,
  {{\partial \left[ \left(\Gamma_3-1\right) \rho\, {\cal V}_j
  \left(\delta q^j\right)_{s^\prime} \right]}
  \over {\partial q^i}}\, dV\! \bigg\}
  \nonumber \\
 & & \sigma_{\lambda^\prime,\nu^\prime}^{-1}\,
  A^{(0)}_{s^\prime}\left(\tilde{t}\,\right)
  \exp \left( {\rm i}\,{\sigma_{\lambda^\prime,\nu^\prime} \over
  \sigma_{\ell,N}} t^+ \right),  \label{b:28b}
\end{eqnarray}
where ${\cal N}_s$ is the square of the norm of the Lagrangian
displacement for the oscillation mode $s$ [Paper~I, Eq.\ (48)].

The integral contained in the second term of the right-hand member of
Eq.\ (\ref{b:28b}) is developed in Eq.\ (44) of Paper~I. After
passing on to dimensionless quantities, one has
\begin{eqnarray}
\lefteqn{ \int_V \rho\, {{\partial W_{\ell,m,k}}\over {\partial q^i}}\,
  \overline{\left({\delta q^i}\right)_s}\, dV
  = - \delta_{\lambda, \ell}\, \delta_{\mu, m}\,
  {{4\, \pi}\over {2\, \ell+1}}\, {{(\ell+|m|)!}\over
  {(\ell - |m|)!}} } \nonumber \\
 & & c_{\ell, m, k} \int_0^1 \ell\, r^{\ell-1}
  \left[{ \xi_{\lambda, \nu}
  + (\ell+1)\, \eta_{\lambda, \nu}/r }\right]
  \rho\, r^2\, dr,  \label{b:29}
\end{eqnarray}
where $\delta_{\lambda, \ell}$ is a Kronecker's
delta, and $\xi_{\lambda,\nu}(r)$ and $\eta_{\lambda,\nu}(r)$ are the
radial parts of the radial and the transverse component of the
Lagrangian displacement [Paper~I, Eqs.\ (43)].

The integral contained in the third term of the right-hand member of
Eq.\ (\ref{b:28b}) can be transformed by means of Gauss' theorem. If
the mass density vanishes at the star's surface, one derives that
\begin{eqnarray}
\lefteqn{\int_V \overline{\left( \delta q^i \right)_s}\,
  \left\{ {\partial \over {\partial q^i}} \left[
  \left(\Gamma_3-1\right) \rho\, {\cal V}_j
  \left(\delta q^j\right)_{s^\prime} \right] \right\} dV} \nonumber \\
 & & =   {{4\, \pi}\over {2\, \lambda+1}}\, {{(\lambda + |\mu|)!}\over
  {(\lambda - |\mu|)!}}\, \delta_{\lambda, \lambda^\prime}\,
  \delta_{\mu, \mu^\prime}  \nonumber \\
 & & \int_0^1 \left(\Gamma_3-1\right)
  \overline{\left( \delta \rho/\rho \right)_{\lambda,\nu}}\,
  \left[ \delta \left( dQ/dt \right)
  \right]_{\lambda^\prime,\nu^\prime}
  \rho\, r^2\, dr.  \label{b:30b}
\end{eqnarray}

In the inhomogeneous part of Eq.\ (\ref{b:28b}), the resonant terms
must be removed. We distinguish between two cases
depending on whether the star's spheroidal mode $s$ is
equal to or different from the spheroidal mode with eigenfrequency
$\sigma_{\ell,N}$ that belongs to the spherical harmonic
$Y_\ell^m(\theta,\phi)$ and is involved in the resonance. We
denote the latter spheroidal mode as the spheroidal mode $S$.

When $s=S$, the resonant terms are removed from the inhomogeneous part
of Eq. (\ref{b:28b}) by setting
\begin{eqnarray}
\lefteqn{\displaystyle {{d A^{(0)}_S}\over {d \tilde{t}}}
  + C^\prime\, {\kappa_{\ell,N} \over \sigma_{\ell,N}}\,
  A^{(0)}_S }  \nonumber \\
 & & = \displaystyle {c_{\ell,m,k}\, Q_{\ell,N} \over {2\, {\rm i} }}\,
  \exp \left[- {\rm i} \left(f\, \tilde{t} + k\, n\, \tau \right)
  \right].  \label{b:33}
\end{eqnarray}
In this equation, the factor $Q_{\ell,N}$ is defined as
\begin{equation}
Q_{\ell, N} =
  {1 \over \sigma_{\ell,N}^2} {{\int_0^{R_1} \ell\, r^{\ell-1}
   \left[ \xi_{\ell,N} + (\ell+1)\, \eta_{\ell,N}/r \right] \rho\,
   r^2\, dr} \over {\int_0^{R_1} \left[ \xi^2_{\ell,N} + \ell(\ell + 1)\,
   \eta^2_{\ell,N}/r^2 \right] \rho\, r^2\, dr }}   \label{Qln}
\end{equation}
[see Paper~I, Eq.\ (50)], and
$\kappa_{\ell,N}$ is the star's coefficient of vibrational
stability for the spheroidal mode $S$ defined as
\begin{equation}
\kappa_{\ell,N} = -
  {{\int_0^1 \left(\Gamma_3-1\right)
  \overline{\left(\delta \rho/\rho \right)_{\ell,N}}\,
  \left[ \delta (dQ/dt) \right]_{\ell,N} \rho\, r^2\, dr } \over
  {2\, \sigma_{\ell,N}^2\,
  \int_0^1 \left[{ \xi^2_{\ell,N} + \ell(\ell + 1)\,
  \eta^2_{\ell,N}/r^2 }\right] \rho\, r^2\, dr}}
  \label{b:35}
\end{equation}
\citep[ Sect.~63]{Led1958}. The star is vibrationally stable
with respect to the spheroidal mode $S$ when $\kappa_{\ell,N}$ is
positive, and vibrationally unstable with respect to that
mode when $\kappa_{\ell,N}$ is negative.

A general solution of Eq.\ (\ref{b:33}) is given by
\begin{eqnarray}
\lefteqn{A^{(0)}_S \left(\tilde{t}\,\right) = a^{(0)}_S\,
  \exp \left(- C^\prime\,
  {\kappa_{\ell,N} \over \sigma_{\ell,N}}\,
  \tilde{t}\, \right)
  + {{\varepsilon_T\, \sigma_{\ell,N}} \over 2} }  \nonumber \\
 & & {{\varepsilon\, \sigma_{\ell,N} - {\rm i}\, C\, \kappa_{\ell,N}}
  \over {\varepsilon^2\, \sigma_{\ell,N}^2 + C^2\, \kappa_{\ell,N}^2}}\,
  c_{\ell,m,k}\, Q_{\ell,N}\,
  \exp \left[ - {\rm i} \left( f\, \tilde{t}
  + k\, n\, \tau \right) \right] \!,  \label{b:36}
\end{eqnarray}
where $a^{(0)}_S$ is an undetermined constant. For the sake of
simplification, we set $a^{(0)}_S = 0$.

Next, when $s \neq S$, one removes the resonant terms from the
inhomogeneous part of Eq.\ (\ref{b:28b}) by setting
\begin{equation}
A^{(0)}_s \left(\tilde{t}\, \right) = 0.  \label{b:38}
\end{equation}
This set of modes $s$ includes the spheroidal modes with
eigenfrequency $\sigma_{\ell,N}$ that belong to the spherical
harmonics $Y_\ell^\mu(\theta,\phi)$ of degree $\ell$ but of
azimuthal number $\mu \ne m$.

By introducing the phase angle $\psi_{k,S}$ of the complex
number $\varepsilon\, \sigma_{\ell,N} - {\rm i}\, C\, \kappa_{\ell,N}$
as
\begin{equation}
\tan \psi_{k,S} = - {{C\, \kappa_{\ell,N}} \over
  {\varepsilon\, \sigma_{\ell,N}}},  \label{b:41}
\end{equation}
it follows that the lowest-order solutions for the components of the
resonant dynamic tide take the form
\begin{eqnarray}
\lefteqn{ \left( \delta q^j \right)_T\left(\vec{r},t\right) =
  {{\varepsilon_T\, c_{\ell,m,k}\, Q_{\ell,N}} \over
  {2 \left( \varepsilon^2
  + C^2\, \kappa_{\ell,N}^2/\sigma_{\ell,N}^2 \right)^{1/2}}}\,
  \left( \delta q^j \right)_{\ell,m,N} \left(\vec{r}\,\right) }  \nonumber \\
 & & \exp
  \left[ \mbox{i} \left( \sigma_T\, t - k\, n\, \tau +
  \psi_{k,S}\right) \right],
  \,\,\, j = 1, 2, 3. \label{b:39f}
\end{eqnarray}

Hence, at the lowest order of approximation in the expansion parameter
$\varepsilon_T$, the solution consists of the resonantly excited
oscillation mode with eigenfrequency $\sigma_{\ell,N}$ that is
associated with the spherical harmonic $Y_\ell^m(\theta,\phi)$.
This conclusion is similar to the conclusion reached
in Paper~I for resonant dynamic tides considered in the isentropic
approximation. However, the amplitude of the resonant dynamic tide is now
proportional to
\[
\varepsilon_T\, c_{\ell,m,k}/\left(\varepsilon^2
  + C^2\, \kappa_{\ell,N}^2/\sigma_{\ell,N}^2 \right)^{1/2}
\]
and is reduced in
comparison to the amplitude reached by a resonant dynamic tide
considered in the isentropic approximation. The reduction results from
the term $C^2\, \kappa_{\ell,N}^2/\sigma_{\ell,N}^2$ in the
denominator and is independent of the sign of the star's coefficient of
vibrational stability for the oscillation mode $S$. A reduction of the
amplitude due to the nonadiabatic effects was already observed by
\citet{Zahn1975} [see the comments below his Eq.\ (2.40)].

A main difference with the solutions found in the isentropic
approximation is that the nonadiabatic effects induce a phase shift
$\psi_{k,S}$ between the tidal motions associated with the resonant
dynamic tide and the tide-generating potential. The resonant dynamic
tide lags behind the tide-generating potential when the star is
vibrationally stable with respect to the oscillation mode $S$, whereas
it precedes the tide-generating potential when the star is
vibrationally unstable with respect to that mode. The
existence of a phase shift due to the nonadiabaticity of the tidal
motions was taken into consideration by Zahn in his earlier
investigation [1975, Eq. (4.5)].  Outside resonance, a phase
shift does not appear until the next order of approximation in the
small expansion parameter \citep{W2000}.

As noted by \citet{Ter1998} in connection with an expression
for the torque caused by dissipation in the convective envelope of a
solar-type star, Solutions (\ref{b:39f}) are valid only for forcing
angular frequencies $\sigma_T$ for which
\begin{equation}
  \varepsilon^2  \gg C^2\, \kappa_{\ell,N}^2/\sigma_{\ell,N}^2,
  \label{vw}
\end{equation}
since the nonadiabatic effects are assumed to be small in the
linear perturbation analysis.

Besides the partial dynamic tide with forcing frequency
$\sigma_T$, we also consider the partial dynamic tide with forcing
frequency $-\sigma_T$. The latter partial dynamic tide is
resonant with the isentropic oscillation mode with eigenfrequency
$-\sigma_{\ell,N}$. One obtains the corresponding solutions for
the components of the tidal displacement by replacing $k$ by $-k$
and $m$ by $-m$ in Solutions (\ref{b:39f}).

When one takes into account the property that $c_{\ell,-m,-k} =
c_{\ell,m,k}$ and reintroduces the dimensions for the various physical
quantities, the global solutions for the components of the resonant
dynamic tide take the form
\begin{equation}
\renewcommand{\arraystretch}{2.2}
\left.
\begin{array}{l c l}
\lefteqn{ (\delta r)_T\left(\vec{r},t\right) =
  {{\varepsilon_T\, c_{\ell,m,k}} \over {\left(\varepsilon^2 +
  \kappa_{\ell,N}^2/\sigma_{\ell,N}^2 \right)^{1/2}}}\,
  Q_{\ell,N} } \nonumber \\
 & &  \xi_{\ell,N}(r)\,
 P_\ell^{|m|}(\cos \theta)\, \cos \beta_{m, k}(\phi,t), \nonumber \\
\lefteqn{ (\delta \theta)_T\left(\vec{r},t\right) =
  {{\varepsilon_T\, c_{\ell,m,k}} \over {\left(\varepsilon^2  +
  \kappa_{\ell,N}^2/\sigma_{\ell,N}^2 \right)^{1/2}}}\,
  Q_{\ell,N} } \nonumber \\
 & & \displaystyle
  {{\eta_{\ell,N}(r)}\over r^2}\,
  {{\partial P_\ell^{|m|}(\cos \theta)} \over
  {\partial \theta}}\, \cos \beta_{m, k}(\phi,t),  \nonumber \\
\lefteqn{ (\delta \phi)_T\left(\vec{r},t\right) = -
  {{\varepsilon_T\, c_{\ell,m,k} } \over {\left(\varepsilon^2  +
  \kappa_{\ell,N}^2/\sigma_{\ell,N}^2 \right)^{1/2}}}\,
  Q_{\ell,N} } \nonumber \\
 & & \displaystyle
 {{\eta_{\ell,N}(r)}\over {r^2 \sin^2 \theta}}\,
  P_\ell^{|m|}(\cos \theta)\, m\,
  \sin \beta_{m, k}(\phi,t),  \hspace{0.5cm}
\end{array}\right\}  \label{b:40}
\end{equation}
where
\begin{equation}
\beta_{m, k}(\phi,t) = m\,\phi + \sigma_T\,t - k\,n\,\tau
  +\psi_{k,S}. \label{eq3}
\end{equation}

Since both the dynamic tide with forcing frequency $\sigma_T$
and the dynamic tide with forcing frequency $-\sigma_T$ are
taken into account in the derivation of Solutions (\ref{b:40}), only
non-negative values of $k$ must be considered subsequent\-ly.

\section{Secular variations of the orbital elements}

The tidal distortion of a star brings about a perturbation of the star's
external gravitational field, which in its turn gives rise to time-dependent
variations of the orbital elements. For nonadiabatic resonant dynamic tides,
the phase shift between the tidal displacement field and the tide-generating
potential induces variations of the semi-major axis and the orbital
eccentricity besides the variation of the longitude of the periastron, which
is already found in the isentropic approximation.

By means of a classical perturbation procedure used in celestial
mechanics, the equations governing the rates of change of the
semi-major axis $a$, the orbital eccentricity $e$, and the longitude
of the periastron $\varpi$ are derived from a perturbing function $R$
as
\begin{equation}
\renewcommand{\arraystretch}{2.2}
\left.
\begin{array}{l c l}
\displaystyle {{da} \over {dt}} & = &
  \displaystyle - {2 \over {n^2\, a}}
  {{\partial R}\over {\partial \tau}},  \nonumber \\
\displaystyle {{de} \over {dt}} & = &
  \displaystyle - {1 \over {n\, a^2\, e}}\!
  \left[{{\left(1-e^2\right)} \over n}
  {{\partial R}\over {\partial \tau}}
  + \left(1-e^2\right)^{1/2} {{\partial R}\over {\partial
  \varpi}} \right] \!, \!  \nonumber \\
\displaystyle {{d\varpi} \over {dt}} & = & \displaystyle
  {1 \over {n\,a^2}}
  {{\left(1 - e^2\right)^{1/2}}\over e}
  {{\partial R}\over {\partial e}}  \nonumber
\end{array}\right\}  \label{sec:4}
\end{equation}
\citep[e.g.,][]{Ste1960}.

The perturbing function $R$ is related to the Eulerian perturbation of
the star's external gravitational potential $\Phi_e^\prime$ at the
instantaneous position of the companion as
\begin{equation}
R(u,v,t) = - {{M_1 + M_2}\over M_1}\, \Phi_e^\prime
   \left({u,{\pi\over 2}, v - \Omega\, t;t}\right),  \label{sec:1}
\end{equation}
where $u$ is the companion's radial distance.

In accordance with the lowest-order Solutions (\ref{b:40}) for the
components of a
nonadiabatic resonant dynamic tide, the Eulerian perturbation of the
gravitational potential at the star's surface takes the form
\begin{eqnarray}
\lefteqn{ \Phi^\prime_T(R_1,\theta,\phi;t) =
  {{\varepsilon_T\, c_{\ell,m,k}} \over {\left(\varepsilon^2  +
  \kappa_{\ell,N}^2/\sigma_{\ell,N}^2 \right)^{1/2}}}\, Q_{\ell,N}\,
  \Phi^\prime_{\ell, N}\left({R_1}\right) } \nonumber \\
 & & P_\ell^{|m|}(\cos \theta)\, \cos \beta_{m, k}(\phi,t).
  \hspace{1cm} \label{b:42}
\end{eqnarray}
Because of the continuity of
the gravitational potential at the star's surface, the Eulerian
perturbation of the external gravitational potential can be expressed
as
\begin{eqnarray}
\lefteqn{ \Phi_e^\prime\left(\vec{r},t\right) =
  {{\varepsilon_T\, c_{\ell,m,k}} \over {\left(\varepsilon^2 +
  \kappa_{\ell,N}^2/\sigma_{\ell,N}^2 \right)^{1/2}}}\,
  Q_{\ell,N}\, \Phi^\prime_{\ell, N}\left({R_1}\right)
  } \nonumber \\
 & & \left({r \over R_1}\right)^{\!\!-(\ell+1)} \!\!
  P_\ell^{|m|}(\cos \theta) \cos \beta_{m, k}(\phi,t).
   \label{sec:2a}
\end{eqnarray}
With the use of Definition (\ref{b:3}) of the small parameter
$\varepsilon_T$, the perturbing function $R$ can then be written as
\begin{eqnarray}
\lefteqn{ R(u,v,t)
  = {{c_{\ell, m, k}\, H_{\ell, N}} \over  {\left(\varepsilon^2 +
  \kappa_{\ell,N}^2/\sigma_{\ell,N}^2 \right)^{1/2}}}\,
  {{G \left( M_1 + M_2 \right)}\over R_1}\, P_\ell^{|m|}(0)
  }  \nonumber \\
 & &  \left({R_1\over a}\right)^{\ell + 4}\! {M_2\over M_1}
  \left({u\over a}\right)^{-(\ell+1)}\! \cos
  \left(m\,v + k\,M + \psi_{k,S}\right).  \label{sec:2}
\end{eqnarray}
Here $H_{\ell, N}$ is a dimensionless factor defined by
Eq. (62) of Paper~I. This
factor renders the combined effect of the overlap integral, which is
proportional to the work done by the tidal force, and the Eulerian
perturbation of the gravitational potential associated with the
resonantly excited oscillation mode.

Substitution of the expression for the perturbing function into Eqs.\
(\ref{sec:4}), transformation of the time derivatives into derivatives
with respect to the mean anomaly $M$, and averaging the derivatives
$da/dM$, $de/dM$, and $d\varpi/dM$ over a revolution of the companion
yield the following equations for the relative rates of secular
change of the orbital elements $a$, $e$, and $\varpi$ resulting from a
nonadiabatic resonant dynamic tide:
\begin{eqnarray}
\lefteqn{{1 \over a}\, \left( {{da} \over {dt}} \right)_{\rm sec}
  = - \left({R_1\over a}\right)^{\ell + 3} {M_2\over M_1}\,
  {{2\, \pi} \over T_{\rm orb}} }  \nonumber \\
 & & {{\sigma_{\ell,N}\, \kappa_{\ell,N}} \over
  {\varepsilon^2\, \sigma_{\ell,N}^2 + \kappa_{\ell,N}^2 }}\,
  H_{\ell, N}\, G_{\ell,m,k}^{(2)}(e),  \label{sec:11}
\end{eqnarray}
\begin{eqnarray}
\lefteqn{{1 \over e}\, \left( {{de} \over {dt}} \right)_{\rm sec}
  = -  \left({R_1\over a}\right)^{\ell + 3} {M_2\over M_1}\,
  {{2\, \pi} \over T_{\rm orb}} }  \nonumber \\
 & & {{\sigma_{\ell,N}\, \kappa_{\ell,N}} \over
  {\varepsilon^2\, \sigma_{\ell,N}^2 + \kappa_{\ell,N}^2 }}\,
  H_{\ell, N}\, {G_{\ell,m,k}^{(3)}(e) \over e},  \label{sec:12}
\end{eqnarray}
\begin{eqnarray}
\lefteqn{{1 \over {2\, \pi}}\,
  \left( {{d \varpi} \over {dt}} \right)_{\rm sec}
  = \left({R_1\over a}\right)^{\ell + 3} {M_2\over M_1}\,
  {1 \over T_{\rm orb}} } \nonumber \\
 & & {{\varepsilon\, \sigma_{\ell,N}^2} \over
  {\varepsilon^2\, \sigma_{\ell,N}^2 + \kappa_{\ell,N}^2 }}\,
  H_{\ell, N}\, G_{\ell,m,k}^{(1)}(e).  \label{sec:10}
\end{eqnarray}
In these equations, $T_{\rm orb}$ is the orbital period, the functions
$G_{\ell,m,k}^{(1)}(e)$ correspond to the functions
$G_{\ell,m,k}(e)$ defined by Expression (71) of Paper~I, and the
functions $G_{\ell,m,k}^{(2)}(e)$ and $G_{\ell,m,k}^{(3)}(e)$ are
defined as
\begin{eqnarray}
\lefteqn{G_{\ell,m,k}^{(2)}(e) = {2 \over {\left(1-e^2\right)^{\ell+1}}}\,
  c_{\ell,m,k}\, P_\ell^{|m|}(0)\, \pi^{-1} } \nonumber \\
 & & \bigg[(\ell+1)\, e\, \int_0^\pi
   (1+e\,\cos v)^{\ell}\, \sin (m\,v + k\,M)\, \sin v\, dv
     \nonumber \\
 & & -\, m\, \int_0^\pi (1+e\,\cos v)^{\ell+1}\,
   \cos (m\,v + k\,M)\, dv \bigg],   \label{sec:14}
\end{eqnarray}
\begin{eqnarray}
\lefteqn{G_{\ell,m,k}^{(3)}(e) = {1 \over {e\, \left(1-e^2\right)^{\ell}}}\,
  c_{\ell,m,k}\, P_\ell^{|m|}(0)\, \pi^{-1} } \nonumber \\
 & & \bigg\{(\ell+1)\, e\, \int_0^\pi
   (1+e\,\cos v)^{\ell}\, \sin (m\,v + k\,M)\, \sin v\, dv
     \nonumber \\
 & & -\, m\, \int_0^\pi (1+e\,\cos v)^{\ell-1}\,
   \left[ (1+e\,\cos v)^2 - \left(1-e^2\right)\right]  \nonumber \\
 & & \cos (m\,v + k\,M)\, dv \bigg\}.   \label{sec:15}
\end{eqnarray}

>From Eqs.\ (\ref{sec:11}) -- (\ref{sec:10}),
it results that the relative rates of secular change of the semi-major
axis, the orbital eccentricity, and the longitude of the periastron
due to a nonadiabatic resonant dynamic tide are proportional
to $\left({R_1/a}\right)^{\ell + 3}$ and to the mass ratio
$M_2/M_1$. These factors correspond to those found for the rate of
secular change of the longitude of the periastron
in the cases of
resonant dynamic tides that are treated in the isentropic
approximation. The same factors also appear in various expressions
derived for the relative rates of secular
change of orbital elements outside any resonance with a free
oscillation mode \citep{Zahn1977,Zahn1978,Hut1981,Sme1991,Ruy1992}.

Secondly, the relative rates of secular change of the semi-major axis,
the orbital eccentricity, and the longitude of the periastron depend on the
resonantly excited oscillation mode through the eigenfrequency
$\sigma_{\ell,N}$, the coefficient of vibrational stability
$\kappa_{\ell,N}$, and the integral expression $H_{\ell,N}$. Two cases
can be distinguished.

For the relative rates of secular change of the semi-major axis and
the orbital eccentricity, the factor related to the resonantly
excited oscillation mode is
\[
  {{\sigma_{\ell,N}\, \kappa_{\ell,N}} \over
  {\varepsilon^2\, \sigma_{\ell,N}^2 + \kappa_{\ell,N}^2}}\, H_{\ell,N}.
\]
The presence of the coefficient of vibrational stability in the
numerator illustrates that the secular changes of the semi-major axis
and the orbital eccentricity result from the nonadiabaticity of the
resonant dynamic tide.

For the relative rate of secular change of the
longitude of the periastron, the factor related to the resonantly
excited oscillation mode is
\[
  {{H_{\ell,N}/ \varepsilon} \over {1 + \kappa_{\ell,N}^2/
   \left( \varepsilon^2\, \sigma_{\ell,N}^2 \right)}}.
\]
The nonadiabaticity of the oscillation mode thus reduces the relative
rate of secular change of the longitude of the periastron in
comparison to the value it takes in the isentropic approximation
[Paper~I, Eq.\ (70)]. The reduction, however, is small because
of Inequality (\ref{vw}).

Thirdly, the relative rates of secular change of the semi-major axis,
the eccentricity, and the longitude of the periastron depend
on the eccentricity through the functions
$G_{\ell,m,k}^{(1)}(e)$, $G_{\ell,m,k}^{(2)}(e)$, and
$G_{\ell,m,k}^{(3)}(e)$, which are characterised by the
degree $\ell$, the azimuthal number $m$, and the number $k$ of
the term considered in the tide-generating potential.

For the rates of secular change of the semi-major axis and the
orbital eccentricity, the order of magnitude in the ratio $R_1/a$ can
be estimated as follows. When one returns to the dimensionless
quantities used in the previous section and expresses the
eigenfrequency $\sigma_{\ell,N}$ and the  coefficient of vibrational
stability $\kappa_{\ell,N}$ in the units $\left( G M_1/R_1^3
\right)^{1/2}$ and $R_1 L_1/\left( G M_1^2\right)$, respectively, the
factor
\[
  \sigma_{\ell,N}\, \kappa_{\ell,N}/ \left(
  \varepsilon^2\, \sigma_{\ell,N}^2 + \kappa_{\ell,N}^2 \right),
\]
which appears in Eqs.\ (\ref{sec:11}) and (\ref{sec:12}), takes the form
\[
  C\, \sigma_{\ell,N}\, \kappa_{\ell,N}/ \left(
  \varepsilon^2\, \sigma_{\ell,N}^2 +
  C^2\, \kappa_{\ell,N}^2 \right).
\]
Because of Condition (\ref{vw}), the approximation holds
\begin{equation}
{{C\, \sigma_{\ell,N}\, \kappa_{\ell,N}} \over
  {\varepsilon^2\, \sigma_{\ell,N}^2 +
  C^2\, \kappa_{\ell,N}^2}} \simeq
  {{C\, \kappa_{\ell,N}} \over {\varepsilon^2\, \sigma_{\ell,N}}}.
  \label{mag1}
\end{equation}
Since both the relative frequency difference $\varepsilon$ and the
nonadiabatic effects are considered
to be of the order of $\varepsilon_T$, the factor is of the
order of $\varepsilon_T^{-1}$ so that
\begin{equation}
{{\sigma_{\ell,N}\, \kappa_{\ell,N}} \over
  {\varepsilon^2\, \sigma_{\ell,N}^2 + \kappa_{\ell,N}^2}}
  \propto \left( {R_1 \over a} \right)^{-3}.  \label{mag2}
\end{equation}
Furthermore, the factor $2\pi/T_{\rm orb}$ can be related to the
ratio $R_1/a$ by means of Kepler's third law as
\begin{equation}
{{2\, \pi} \over T_{\rm orb}} = \left( {R_1 \over a} \right)^{3/2}
  \left( {{G\, M_1} \over R_1^3} \right)^{1/2}
  \left( {{M_1 + M_2} \over M_1} \right)^{1/2}.  \label{mag3}
\end{equation}

It follows that
\begin{equation}
{1 \over a} \left( {{da} \over {dt}} \right)_{\rm sec}\!\!
  \propto \left({R_1\over a}\right)^{\ell + 3/2}\!\!, \,\,\,\,
{1 \over e} \left( {{de} \over {dt}} \right)_{\rm sec}\!\!
  \propto \left({R_1\over a}\right)^{\ell + 3/2}\!\!\!.
  \label{sec:12b}
\end{equation}
In particular, for $\ell=2$, the relative rates of secular change of
the semi-major axis and the orbital eccentricity are of the order of
$\left( R_1/a \right)^{7/2}$. This relative rate of secular change of
the orbital eccentricity is considerably larger than that found by
\citet{Zahn1977} for the limiting case of small forcing
frequencies: in that case, the relative rate of secular change of the
orbital eccentricity is of the order $\left( R_1/a \right)^{21/2}$.

Finally, it may be observed
that the relative rates of secular change of the semi-major axis and
the orbital eccentricity due to a nonadiabatic resonant dynamic tide
may vary considerably from one mode to another by the value of the
factor $H_{\ell,N}$.

\section{Secular change of the star's rotational velocity}

As well because of the phase shift induced by a nonadiabatic resonant
dynamic tide, the companion exerts a torque $\vec{\cal T}$ on the
tidally  distorted star, which is normal to the orbital plane.
This torque can be derived from Newton's law of action and reaction as
the opposite of the torque exerted by the tidally
distorted star on the companion due to the perturbation of the star's
external gravitational field:
\begin{equation}
\vec{\cal T} = M_2 \left({\vec{r} \times \nabla
\Phi_e^\prime}\right) \label{torque:3}
\end{equation}
\citep{Zahn1975}.

By the use of Solution (\ref{sec:2a}) for the Eulerian perturbation of
the external gravitational potential, the torque is given by
\begin{eqnarray}
\lefteqn{{\cal T} = - {{\varepsilon_T\, c_{\ell,m,k}}
  \over {\left(\varepsilon^2 +
  \kappa_{\ell,N}^2/\sigma_{\ell,N}^2 \right)^{1/2}}}\, Q_{\ell,N}\,
  \Phi^\prime_{\ell, N}\left({R_1}\right)
  \left( {R_1 \over a} \right)^{\ell+1} }  \nonumber \\
 & & \displaystyle
  M_2\, P_\ell^{|m|}(0)\, m\,  \left( {u \over a} \right)^{-(\ell+1)}
  \sin \left( m\,v + k\,M + \psi_{k,S} \right).
  \hspace{0.5cm} \label{ang:3}
\end{eqnarray}
After averaging over a revolution of the companion, one obtains
\begin{eqnarray}
\lefteqn{ \overline{\cal T} = -
  \left({R_1\over a}\right)^{\ell + 3} {M_2\over M_1}\,
  \left( {{G\, M_1^2\, M_2^2} \over {M_1+M_2}} \right)^{1/2}
  {{2\, \pi} \over T_{\rm orb}}\, a^{1/2}
  }  \nonumber \\
 & &   {{\sigma_{\ell,N}\, \kappa_{\ell,N}} \over
  {\varepsilon^2\, \sigma_{\ell,N}^2 + \kappa_{\ell,N}^2 }}\,
  H_{\ell, N}\, G^{(4)}_{\ell,m,k}(e),  \label{ang:4bb}
\end{eqnarray}
where $G^{(4)}_{\ell,m,k}(e)$ is a function of the orbital eccentricity
defined as
\begin{equation}
G^{(4)}_{\ell,m,k}(e) = m\, {{(\ell+|m|)!} \over {(\ell-|m|)!}}\,
  \left( {R_1 \over a} \right)^{-(\ell-2)} c_{\ell,m,k}^2.  \label{ang:5}
\end{equation}
The function $G^{(4)}_{\ell,m,k}(e)$ is different from zero only for
nonaxisymmetric, nonadiabatic resonant dynamic tides and has the same
sign as the azimuthal number $m$. It is related to the
functions $G^{(2)}_{\ell,m,k}(e)$ and $G^{(3)}_{\ell,m,k}(e)$ as
\begin{equation}
G^{(4)}_{\ell,m,k}(e) = {e \over {\left(1-e^2\right)^{1/2}}} \!
  \left[ G^{(3)}_{\ell,m,k}(e)\! -\! {{1-e^2} \over {2\,e}}
  G^{(2)}_{\ell,m,k}(e)\! \right]\!\!.  \label{ang:5c}
\end{equation}

The tidal torque $\vec{\cal T}$ exerted by the companion on the
tidally distorted star affects the angular velocity of the star's
rotation in accordance with the conservation of total angular
momentum of the binary
\begin{equation}
\left({{d {\cal L}_{\rm orb}} \over {dt}} \right)_{\rm sec} +
  \left({{d {\cal L}_{\rm rot}} \over {dt}} \right)_{\rm sec} = 0,
  \label{ang:6}
\end{equation}
where the orbital angular momentum ${\cal L}_{\rm orb}$ is determined
as
\begin{equation}
{\cal L}_{\rm orb} = \left( {{G\, M_1^2\, M_2^2} \over {M_1+M_2}}
  \right)^{1/2} a^{1/2}\, \left( 1-e^2 \right)^{1/2}  \label{torq:7}
\end{equation}
\citep[see, e.g.,][]{Alex1973}. Differentiation with respect to time
and use of Eqs.\ (\ref{sec:11}), (\ref{sec:12}), and (\ref{ang:6})
yield for the rate of secular change of rotational angular
momentum
\begin{eqnarray}
\lefteqn{
  \left({{d {\cal L}_{\rm rot}} \over {dt}} \right)_{\rm sec} =
  - \left({R_1\over a}\right)^{\ell + 3} {M_2\over M_1}\,
  \left( {{G\, M_1^2\, M_2^2} \over {M_1+M_2}} \right)^{1/2}
   } \nonumber \\
 & & {{2\, \pi} \over T_{\rm orb}}\, a^{1/2}\,
  {{\sigma_{\ell,N}\, \kappa_{\ell,N}} \over
  {\varepsilon^2\, \sigma_{\ell,N}^2 + \kappa_{\ell,N}^2 }}\,
  H_{\ell, N}\,  G^{(4)}_{\ell,m,k}(e).
  \label{Lrot:1}
\end{eqnarray}
Hence, angular momentum is exchanged between the star's rotation and
the orbital motion for {\it nonaxisymmetric}, non\-adiabatic
resonant dynamic tides.

Under the assumption that the star rotates as a rigid body, the
rotational angular momentum is determined as
\begin{equation}
{\cal L}_{\rm rot} = I\, \Omega,  \label{torq:5}
\end{equation}
where $I$ is the star's moment of inertia
with respect to the rotation axis. Since, at the lowest-order of
approximation, the rate of secular change of the star's moment
of inertia due to a resonant dynamic tide is equal to zero,
differentiation of Equality (\ref{torq:5}) with respect to time and
use of Eq. (\ref{Lrot:1}) yield for the relative rate of secular
change of the star's angular velocity
\begin{eqnarray}
\lefteqn{ {1 \over \Omega}
  \left({{d \Omega} \over {dt}} \right)_{\rm sec} =
  - \left({R_1\over a}\right)^{\ell + 3} {M_2\over M_1}\,
  \left( {{G\, M_1^2\, M_2^2} \over {M_1+M_2}} \right)^{1/2}
   } \nonumber \\
 & & {{2\, \pi} \over T_{\rm orb}}\, {a^{1/2} \over {I\, \Omega}}\,
  {{\sigma_{\ell,N}\, \kappa_{\ell,N}} \over
  {\varepsilon^2\, \sigma_{\ell,N}^2 + \kappa_{\ell,N}^2 }}\,
  H_{\ell, N}\,  G^{(4)}_{\ell,m,k}(e).
  \label{ang:4}
\end{eqnarray}

One derives the order of magnitude in the ratio $R_1/a$ for the
relative rate of secular change of the star's angular velocity
by proceeding as for the relative
rates of secular change of the semi-major axis and the orbital
eccentricity. By expressing the eigenfrequency $\sigma_{\ell,N}$, the
coefficient of vibrational stability $\kappa_{\ell,N}$, and the star's
moment of inertia $I$ respectively in the units $\left( G M_1/R_1^3
\right)^{1/2}$, $R_1 L_1/\left( G M_1^2\right)$, and $M_1 R_1^2$,
setting $\Omega = \gamma\, n$, with $\gamma \in \bbbr$, and using
Kepler's third law, it follows that
\begin{equation}
{1 \over \Omega} \left({{d \Omega} \over {dt}} \right)_{\rm sec}\!
  \propto \left( {R_1 \over a} \right)^{\ell-1/2}\!.
\end{equation}
Hence, for $\ell=2$, the relative rate of secular change of
the angular velocity is of the order of
$\left(R_1/a\right)^{3/2}$, which is considerably larger than
the order of $\left( R_1/a \right)^{17/2}$ found by \citet{Zahn1977}
for the limiting case of small forcing frequencies.

With regard to synchronisations of components of close binaries, an
appropriate quantity is the relative rate of secular change of the
difference between the star's angular velocity and the companion's
mean motion:
\begin{eqnarray}
\lefteqn{{1 \over {\Omega - n}} \left[ {{d ( \Omega - n )} \over
  {dt}} \right]_{\rm sec} = - \left({R_1\over a}\right)^{\ell + 3}
  {M_2\over M_1}\, {{2\, \pi} \over T_{\rm orb}} }  \nonumber \\
 & & {{\sigma_{\ell,N}\, \kappa_{\ell,N}} \over
  {\varepsilon^2\, \sigma_{\ell,N}^2 + \kappa_{\ell,N}^2 }}\,
  H_{\ell, N}\, {1 \over {\Omega-n}}  \nonumber \\
 & & \left[ {{G\, M_1\, M_2} \over {n\, a\, I_3}}\, G^{(4)}_{\ell,m,k}(e)
  + {3 \over 2}\, n\, G^{(2)}_{\ell,m,k}(e) \right]. \label{ang:8}
\end{eqnarray}

\section{Nonadiabatic tidal resonances in a $5\,M_\odot$ zero-age main
  sequence star}

We have determined phase shifts $\psi_{k,S}$ for nonadiabatic
dynamic tides in resonance with lower-order second-degree $g^+$-modes
in a $5\,M_\odot$ zero-age main sequence stellar model. The model
consists of a convec\-tive core and a radiative envelope and
has a central hydrogen abundance $X_c=0.7$ and a radius
$R_1=2.8\,R_\odot$. The dynamic time scale of the model is 54.8
minutes, the Helmholtz-Kelvin time scale $4.88 \times 10^5$ years, and
the nuclear  time scale $8.65 \times 10^7$ years. The ratio $C$ of the
dynamic time scale to the Helmholtz-Kelvin time scale is equal to
$2.14 \times 10^{-10}$.

We have considered resonances with the second-degree
$g^+$-modes of radial orders $N=1,2,\ldots,20$. The eigenfrequencies
$\sigma_{2,N}$ of these modes are given in Table~\ref{puls1}.
For the determination of the coefficient of vibrational stability
within the framework of the quasi-isentrop\-ic approximation, we
adopted the expressions given by \citet{Deg1992}. Since the
quasi-isentrop\-ic approximation breaks down near the star's surface,
the integration in the numerator of Definition (\ref{b:35}) for the
coefficient of vibrational stability is often stopped at the
transition between the quasi-isentropic stellar interior and the
nonadiabatic surface layers \citep[see, e.g.,][]{Sto1970,Cox1980}. We
therefore determined the numerator in Definition (\ref{b:35}) by
breaking off the integration at various radial distances close to the
star's surface and found the value of the integral to be rather
insensitive to the radial distance at which the integration was
stopped. However, the resulting values of the coefficient of vibrational
stability $\kappa_{2,N}$ differed strongly from, or were even in
contradiction with results of common nonadiabatic stability analyses of slowly
pulsating B stars \citep[see, e.g.,][]{Dzi1993,GS1993}. Several modes
that were expected to be unstable turned out to be stable within the
framework of the quasi-isentropic approximation, and vice
versa. Therefore, we have not used the coefficients of
vibrational stability determined by means of the quasi-isentropic
approximation but adopted the linear growth rates, which are the
opposites of the imaginary parts of the complex eigenfrequencies of
the nonadiabatic modes depending on time by $\exp[{\rm
  i}(\sigma\!+\!{\rm i}\kappa)t]$. This change is justified by the
fact that a first-order perturbation analysis of resonant dynamic
tides within the framework of a fully nonadiabatic treatment yields
solutions for these tides that are formally identical to Solutions
(\ref{b:39f}), except that the linear growth rate of the mode appears
instead of the coefficient of vibrational stability, and that the
factor $Q_{\ell,N}$ is determined in terms of the nonadiabatic
eigenfunctions (see the Appendix). Since, for main-sequence stars, the
real parts of the nonadiabatic eigenfunctions are known to be much
larger than the imaginary parts and almost equal to the isentropic
eigenfunctions \citep[see, e.g.,][]{Unno1989}, the nonadiabatic factor
$Q_{\ell,N}$ has been approximated by means of the expression given by
Equality (\ref{Qln}).

The linear growth rates $\kappa_{2,N}$ for the $5\,M_\odot$ ZAMS
stellar model are presented in Table~\ref{puls1}. The model is
vibrationally stable for all $g^+$-modes considered, except for the
modes $g^+_{10}$ -- $g^+_{14}$.

\begin{table}
\caption{The eigenfrequencies and the linear growth rates for the
  second-degree modes $g_1^+$ -- $g_{20}^+$ of the $5\,M_\odot$ ZAMS
  model. \label{puls1}}
\begin{tabular}{ccc}
\hline \vspace{-0.35cm} \\ mode & $\sigma_{2,N}\, (s^{-1})$ &
$\kappa_{2,N}\, (s^{-1})$
\vspace{0.02cm} \\ \hline \vspace{-0.35cm} \\
$g_1^+$ & $0.661 \; 10^{-3}$ & $0.372 \; 10^{-7}$ \\
$g_2^+$ & $0.460 \; 10^{-3}$ & $0.127 \; 10^{-7}$ \\
$g_3^+$ & $0.350 \; 10^{-3}$ & $0.552 \; 10^{-7}$ \\
$g_4^+$ & $0.281 \; 10^{-3}$ & $0.533 \; 10^{-7}$ \\
$g_5^+$ & $0.234 \; 10^{-3}$ & $0.325 \; 10^{-10}$ \\
$g_6^+$ & $0.200 \; 10^{-3}$ & $0.315 \; 10^{-11}$ \\
$g_7^+$ & $0.175 \; 10^{-3}$ & $0.168 \; 10^{-10}$ \\
$g_8^+$ & $0.156 \; 10^{-3}$ & $0.182 \; 10^{-10}$ \\
$g_9^+$ & $0.141 \; 10^{-3}$ & $0.224 \; 10^{-11}$ \\
$g_{10}^+$ & $0.129 \; 10^{-3}$ & $-0.350 \; 10^{-10}$ \\
$g_{11}^+$ & $0.119 \; 10^{-3}$ & $-0.126 \; 10^{-9}$ \\
$g_{12}^+$ & $0.110 \; 10^{-3}$ & $-0.220 \; 10^{-9}$ \\
$g_{13}^+$ & $0.102 \; 10^{-3}$ & $-0.271 \; 10^{-9}$ \\
$g_{14}^+$ & $0.955 \; 10^{-4}$ & $-0.117 \; 10^{-9}$ \\
$g_{15}^+$ & $0.897 \; 10^{-4}$ & $0.489 \; 10^{-9}$ \\
$g_{16}^+$ & $0.846 \; 10^{-4}$ & $0.175 \; 10^{-8}$ \\
$g_{17}^+$ & $0.800 \; 10^{-4}$ & $0.391 \; 10^{-8}$ \\
$g_{18}^+$ & $0.758 \; 10^{-4}$ & $0.741 \; 10^{-8}$ \\
$g_{19}^+$ & $0.720 \; 10^{-4}$ & $0.127 \; 10^{-7}$ \\
$g_{20}^+$ & $0.686 \; 10^{-4}$ & $0.201 \; 10^{-7}$\vspace{0.05cm}
\\
\hline
\end{tabular}
\end{table}

We have determined the phase shifts $\psi_{k,S}$ on the
supposition that $\varepsilon = 0.001$. The resulting values
are displayed in Fig.~\ref{phase} as a function of the radial order of
the mode. Vibrationally stable modes are represented by solid circles,
and vibrationally unstable modes by open circles.

\begin{figure}
\begin{center}
\resizebox{\hsize}{8cm}{\includegraphics{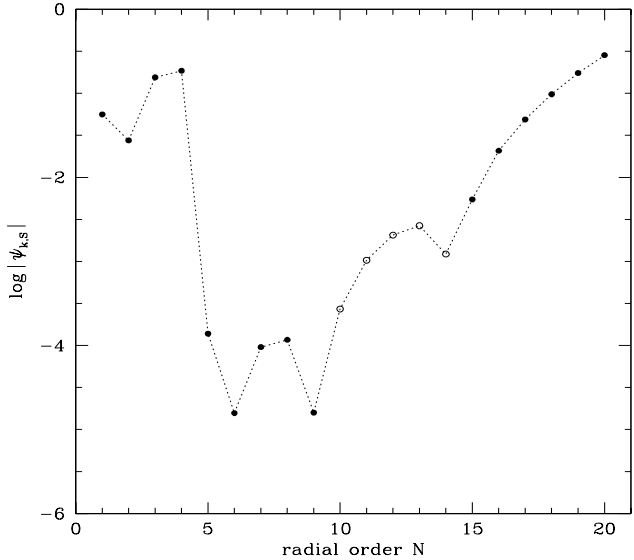}}
\caption{Variation of the logarithm of the phase shift
$\psi_{k,S}$ as a function of the radial order $N$ of the mode for
the $5\,M_\odot$ ZAMS stellar model, and for $\varepsilon=0.001$.
Vibrationally stable modes are represented by solid circles, and
vibrationally unstable modes by open circles. }
\label{phase}
\end{center}
\end{figure}

In addition to the phase shifts, we have determined the
characteristic time scales $t_a$, $t_e$, $t_\Omega$, and
$t_{\rm syn}$ associated with the relative rates of secular change of
the semi-major axis, the orbital eccentricity, the angular
velocity, and the difference between the star's angular velocity and
the companion's mean motion as
\[
  { 1 \over t_a} = \left| {1 \over a}\,
  \left( {{da} \over {dt}} \right)_{\rm sec} \right|, \hspace{0.8cm}
  {1 \over t_e} = \left| {1 \over e}\,
  \left( {{de} \over {dt}} \right)_{\rm sec} \right|,  \label{te}
\]
\[
  {1 \over t_\Omega} = \left| {1 \over \Omega}
  \left( {{d\Omega} \over {dt}} \right)_{\rm sec} \right|, \hspace{0.5cm}
  {1 \over t_{\rm syn}} = \left| {1 \over {\Omega - n}}
  \left[ {{d ( \Omega - n )} \over {dt}} \right]_{\rm sec} \right|.
  \label{ts}
\]
For this, we have set $M_2=1.4\,M_\odot$ and have adopted a value for
the star's rotational angular velocity equal
to $50\%$ of the companion's orbital angular velocity at the
periastron. Furthermore, we have considered the relatively large
orbital eccentricity $e = 0.5$.

In the determination of the time scales, we have taken into
consideration a large number of terms in the expansion of the
tide-generating potential. We have also taken account of the
limitations imposed by the perturbation theory: since
$\varepsilon$ is assumed to be of the order of $\varepsilon_T$ and
must satisfy Inequality (\ref{vw}), we have restricted the values of
$\varepsilon$ to the range of values extending from $\varepsilon =
0.1\, \varepsilon_T$ to $\varepsilon = 10\, \varepsilon_T$.

The contributions of the nonadiabatic resonant dynamic tides to the
secular variations of the orbital elements and of the star's angular
velocity have been restricted to the dominant contributions,
which are those associated with the azimuthal number $m = - 2$.

For orbital periods ranging from 2 to 5 days, the variations of the
time scales, expressed in years, are displayed in Fig.\ \ref{time} on
a logarithmic scale. The gaps in the curves appear at orbital periods
at which the relative frequency difference $\varepsilon$ is outside
the admitted range of values. The black lines correspond to orbital
periods at which $da/dt<0$, $de/dt<0$, $d\Omega/dt>0$, and
$[1/(\Omega-n)][d(\Omega-n)/dt]<0$, and the grey lines to
orbital periods at which $da/dt>0$, $de/dt>0$, $d\Omega/dt<0$, and
$[1/(\Omega-n)][d(\Omega-n)/dt]>0$. The dotted horizontal line
represents the logarithm of the star's nuclear time scale.

The most striking result is that numerous resonances occur which have
strong effects on the relative rates of secular change of the orbital
elements and the star's rotational angular velocity. For the shorter
orbital periods, many resonances lead to characteristic time scales
that are considerably shorter than the star's nuclear time scale.

The resonantly excited oscillation modes in the considered range of
orbital periods are the $g^+$-modes of radial orders $N = 1,
\ldots,12$. At several orbital periods, even resonances with two or
more oscillation modes occur. The feature near the orbital
period of 3.25 days, for instance, is caused by resonances with the
modes $g^+_2$, $g^+_3$, and $g^+_4$.  Furthermore, resonances with
vibrationally unstable modes contribute to secular changes of the
orbital elements and of the star's angular velocity in the sense
opposite to that of the contributions of the resonances with
vibrationally stable modes, so that they counteract the effects
of the latter resonances. At the orbital period of 3.66
days, for instance, the resonance with the mode $g^+_{10}$
contributes to secular changes of the orbital elements and of the
star's angular velocity in the sense opposite to that of the
contributions stemming from the resonances with the modes
$g^+_3$ and $g^+_4$.

\begin{figure*}
\begin{center}
\resizebox{\hsize}{!}{\rotatebox{270}{\includegraphics{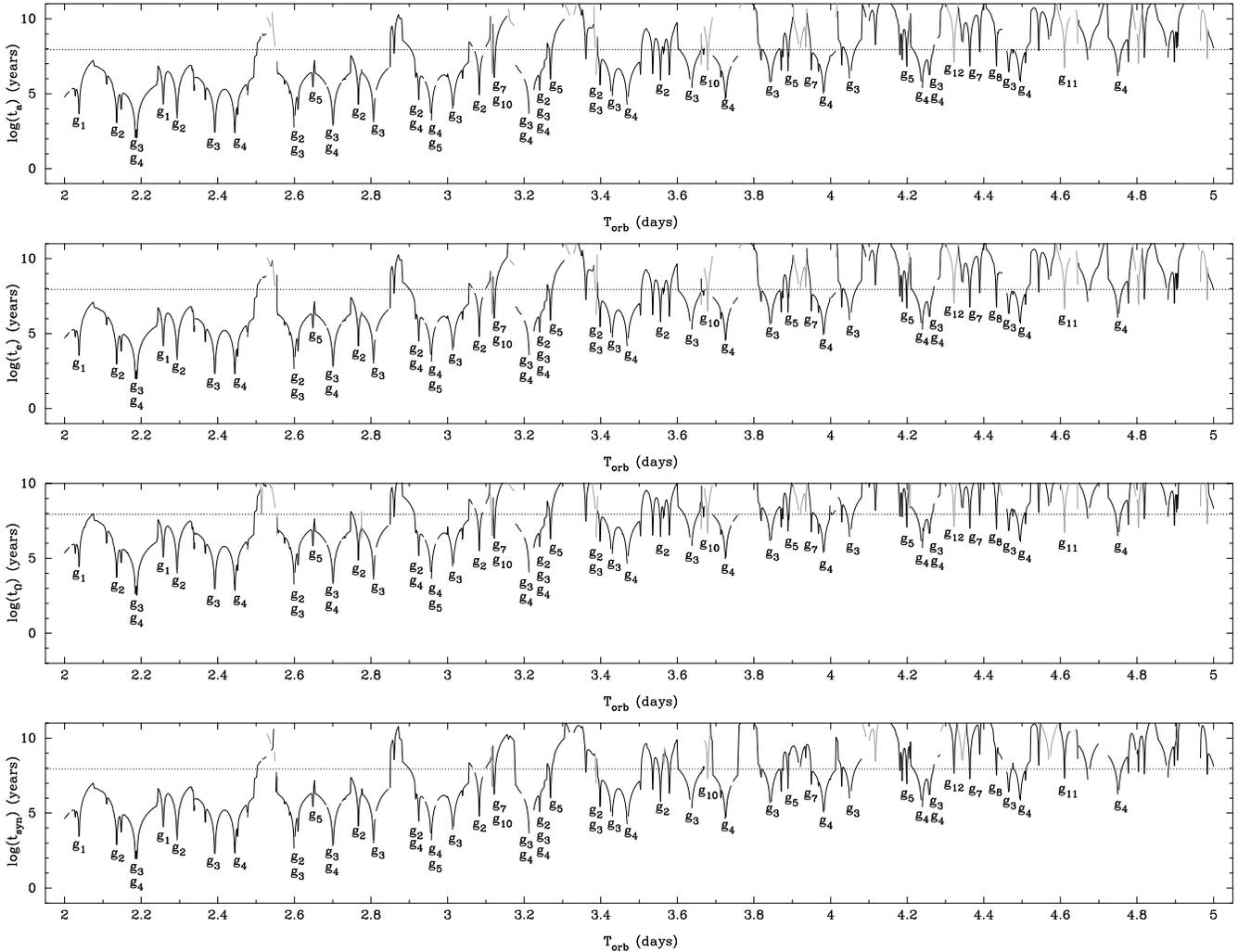}}}
\caption{The logarithms of the time scales $t_a$, $t_e$, $t_\Omega$,
  and $t_{\rm syn}$ as functions of the orbital period for the
  $5\,M_\odot$ ZAMS model and for an orbital eccentricity $e=0.5$. The
  black lines correspond to orbital periods at which $da/dt<0$,
  $de/dt<0$, $d\Omega/dt>0$, and $[1/(\Omega-n)][d(\Omega-n)/dt]<0$,
  and the grey lines to orbital periods at which $da/dt>0$,
  $de/dt>0$, $d\Omega/dt<0$, and
  $[1/(\Omega-n)][d(\Omega-n)/dt]>0$. The dotted horizontal line
  represents the logarithm of the star's nuclear time scale.}
\label{time}
\end{center}
\end{figure*}

With the exception of the shortest orbital periods considered, the
time scales for the relative rates of secular change of the orbital
elements and the star's rotational angular velocity are mostly longer
than the star's nuclear time scale, in accordance with the results of
earlier investigations
\citep[e.g.,][]{SP1983}. The shortest time scales found
for the relative rates of secular change are about five
orders of magnitude smaller than the nuclear time scale.
With regard to the synchronisation of the star's rotation with the
orbital motion of the companion, the time scales found are generally
shorter than the time scales associated with the relative rates of
secular change of the orbital
eccentricity. Synchronisation of the star's rotation with the orbital
motion of the companion at the periastron of its relative orbit can
therefore be expected to occur before the circularisation of the
orbit. A similar conclusion was reached by \citet{Zahn1977} for
low-frequency dynamic tides, after averaging over the effects of
resonances.

When the orbital period becomes larger, nonadiabatic resonant
dynamic tides have a decreasing influence on the time scales for the
relative rates of secular change of the orbital elements and the
star's angular velocity. The underlying reason
is the rapid decrease of the quantity $H_{2,N}$ for higher-order
$g^+$-modes, which are the relevant modes for resonances of dynamic
tides in close binaries with longer orbital periods (see also Paper~I).

On the basis of these results, we suggest that nonadiabatic
resonant dynamic tides may be an efficient mechanism
for the circularisation of orbits in close X-ray binaries, especially
when the orbital period is short. X-ray binaries are thought to have
initially highly eccentric orbits as a result of the supernova
explosion which generates the compact star. However, the orbits of
several X-ray binaries with short orbital periods have been reported
to be circular. Two examples are the X-ray binaries Cen~X-3 and
SMC~X-1. The orbital period of the system Cen~X-3 is 2.1 days, and
that of the system SMC~X-1 3.9 days \citep{Ver1995}. For close binaries
with such short orbital periods and highly
eccentric orbits, resonances of dynamic tides with low-order
$g^+$-modes are quite possible.

\section{Concluding remarks}

We have studied the effects of nonadiabatic resonant dynamic
tides in a component of a close binary on the orbital elements and on
the star's axial rotation. The companion is considered to move in a
fixed Keplerian orbit, and the star is supposed to rotate uniformly
around an axis perpendicular to the orbital plane,
but the effects of the centrifugal force and the Coriolis force
are neglected. The relative difference between the frequency
imposed by the companion and the star's eigenfrequency is assumed
to be of the order of the ratio of the tidal force to the gravity
at the star's equator. Furthermore, the nonadiabatic effects are
considered to be of the same order of magnitude at least in some
region near the star's surface.

We have derived semi-analytical solutions for the linear,
nonadiabatic resonant dynamic tides by applying a two-time variable
expansion procedure, in which the ratio of the tidal force to the
gravity at the star's equator is adopted as the small expansion
parameter. In this perturbation method, the nonadiabatic effects are
evaluated in the quasi-isentropic approximation.
At the lowest order of approximation, the resonant dynamic tide
corresponds to the free oscillation mode involved in the
resonance. This conclusion is similar to the conclusion reached by
\citet{SWV1998} for resonant dynamic tides treated in the isentropic
approximation.

Due to the nonadiabatic effects, the amplitude of the resonantly
excited oscillation mode is reduced in comparison to the amplitude
it reaches in the isentropic approximation. The magnitude of the
reduction is determined by the ratio of the coefficient of
vibrational stability to the eigenfrequency of the oscillation
mode.

The main difference with the dominant solutions derived in the
isentropic approximation is that the nonadiabatic effects induce a
{\it phase shift} between the tidal displacement and the
tide-generating potential. The magnitude and the sign of the phase
shift depend on the coefficient of vibrational stability and on
the relative difference between the forcing angular frequency and
the eigenfrequency of the oscillation mode involved in the
resonance. The tidal displacement field lags behind the
tide-generating potential when the star is vibrationally stable
with respect to the oscillation mode, whereas it precedes the
tide-generating potential when the star is vibrationally unstable
with respect to the oscillation mode.

The phase shift leads to variations of the
semi-major axis and the eccentricity of the companion's relative
orbit, in addition to the variation of the longitude of the
periastron already found in the isentropic approximation of a
resonant dynamic tide. The contributions to the relative rates of
secular change of the semi-major axis and the orbital eccentricity
are determined by Eqs.\ (\ref{sec:11}) and (\ref{sec:12}) and are
of the order of
$\left(R_1/a\right)^{7/2}$, which is much larger than the order
of $\left(R_1/a\right)^{21/2}$ derived by \citet{Zahn1977} for the
limiting case of dynamic tides with small frequencies.

As well because of the phase shift, the companion exerts
a torque on the tidally distorted star, which is normal to the
orbital plane. The average value of the torque is given by Eq.\
(\ref{ang:4bb}). For nonaxisymmetric resonant dynamic tides,
angular momentum is exchanged between the orbital
motion and the star's rotation.
On the assumption that the star rotates as a
rigid body, the relative rate of secular change of the
angular velocity due to a resonant dynamic tide is determined by
Eq.\ (\ref{ang:4}) and is of the order of $\left(R_1/a\right)^{3/2}$,
which is also considerably larger than the order of
$\left(R_1/a\right)^{17/2}$ derived by \citet{Zahn1977} for the
limiting case of dynamic tides with small frequencies.

We have applied our results to resonances of dynamic tides with
second-degree, lower-order $g^+$-modes in a $5\, M_\odot$ zero-age
main sequence stellar model consisting of a convective core and a
radiative envelope. For the orbital eccentricity $e = 0.5$, numerous
resonances of dynamic tides with $g^+$-modes may occur in the range of
orbital periods from 2 to 5 days. At several orbital periods,
even resonances with two or more $g^+$-modes may take place.

Since the quasi-isentropic approximation led to inadequate
results for the coefficient of vibrational stability, we passed on
to the linear growth rates determined within the framework of a fully
nonadiabatic analysis.


Relative to the resonances, we have determined time scales
associated with the rates of secular change of the semi-major axis,
the orbital eccentricity, and the star's rotational angular
velocity. The shortest time scales are found for the shortest orbital
periods and for resonances with the lowest-order $g^+$-modes. They
are about five
orders of magnitude smaller than the star's nuclear time scale.
The time scales for synchronisation of the star's rotation with
the companion's orbital motion at the periastron in the relative
orbit are generally shorter than the time scales for
circularisation of the orbit.

Because of the shortness of the time scales, the system may be
expected to rapidly evolve away from any resonance between a dynamic
tide and a free oscillation mode, so that only a
limited amount of angular momentum can be exchanged between the star's
rotation and the orbital motion of the companion. However,
\citet{Wit1999b,Wit2001} have shown that when both stellar and orbital
evolution are taken into account, a dynamic tide can easily become
locked in a resonance for a prolonged period of time. In these
circumstances, the exchange of angular momentum might be sustained
throughout the locking of the resonance.

For conclusion, in components of close binaries with short orbital
periods and highly eccentric orbits, various resonant dynamic
tides can be excited which, through their nonadiabatic character,
may act as an efficient mechanism for the synchronisation of
the rotation and the circularisation of the orbital motion of stellar
components in time scales shorter than their nuclear time scales.

\begin{acknowledgements}
The authors express their sincere thanks to Dr.\ A. Claret for providing
them with a $5\,M_\odot$ ZAMS stellar model and to Dr.\ A. Gautschy
for allowing them to use his nonadiabatic oscillation code based on the
Riccati method. They also thank an anonymous referee whose critical
comments have led to a substantial improvement of the paper. BW thanks the
Fund for Scientific Research - Flanders (Belgium) for the fellowship of
Research Assistant.
\end{acknowledgements}

\appendix

\section{Resonant dynamic tides within a fully nonadiabatic framework}

In this appendix, we briefly describe how the lowest-order
solution for a nonadiabatic resonant dynamic tide can be derived
within a fully nonadiabatic framework.

Following \citet{TVH1994}, we express the equations that govern
linear, nonadiabatic stellar oscillations in the form of a
first-order, linear differential equation in the vector
${\vec z}=(\vec{\xi},\partial\vec{\xi}/\partial t,\delta S)$, where
$\vec{\xi}$ is the Lagrangian displacement, and $\delta S$ the
Lagrangian perturbation of the entropy. Adding the tidal force exerted
by a companion star, one has
\begin{equation}
{\partial {\bf z}\over\partial t}=M{\bf z}+\vec{N},
\end{equation}
where $\vec{N}=(0,-\varepsilon_T \nabla W,0)$, and $M$ is the
integro-differential operator governing free nonadiabatic stellar
oscillations as defined in
\citet{TVH1994}. The vector ${\bf z}$ is expanded in terms of linear,
nonadiabatic eigenvectors ${\bf e}_n$ as
\begin{equation}
{\bf z}(t,{\bf r})=\sum_na_n(t){\bf e}_n({\bf r}),
\end{equation}
where the eigenvectors ${\bf e}_n$ depend on time by $\exp(\sigma_nt)$
and obey the equation
\begin{equation}
M{\bf e}_n=\sigma_n{\bf e}_n.
\end{equation}
The complex angular frequency $\sigma_n$ can be expressed as
\begin{equation}
\sigma_n=\kappa_n+{\rm i}\Omega_n,
\end{equation}
where $\Omega_n$ is the oscillation frequency, and $\kappa_n$
the opposite of the linear growth rate.

In order to derive a first-order expression for resonant dynamic
tides, we use a multiple time scales method \citep{Nay1973,
Nay1981}. Therefore, we
introduce time variables $T_0$ and $T_1$ as
\begin{equation}
T_0=t, \,\,\, T_1=\varepsilon_T t,
\end{equation}
and expand the amplitude $a_S(t)$ of the resonant mode in terms of the
small parameter $\varepsilon_T$ as
\begin{equation}
a_S=a_{S}^{(0)}+\varepsilon_T a_{S}^{(1)}.
\end{equation}
In addition, we assume the linear growth rate to be small and of
the order of $\varepsilon_T$. In contrast with the procedure
followed in the quasi-isentropic treatment, we here do not consider
the nonadiabatic effects to be small everywhere in the star.

The application of the method of the multiple time scales as
described in \citet{TVH1994} leads to an equation for the first-order
real part of the eigenfunction that has the same form as Eq.\
(\ref{b:39f}), except for the fact that $\kappa_{\ell,N}$ now stands
for the linear growth rate, and the fact that the factor $Q_{\ell,N}$
is defined in terms of the complex nonadiabatic eigenfunctions of the mode.

\aareferences

\end{document}